\pdfoutput=1
\documentclass[a4paper,11pt]{article}
\usepackage{etoolbox}
\newtoggle{blackandwhite}
\togglefalse{blackandwhite} 
\usepackage{authblk}

\usepackage{amsmath}
\usepackage{amssymb}
\usepackage{color}
\usepackage{float}
\usepackage{graphicx}
\usepackage{hhline}
\usepackage{hyperref}
\usepackage{lscape}
\usepackage[pagewise]{lineno}
\usepackage{psfrag}
\usepackage{rotating}
\usepackage{siunitx}
\usepackage{subcaption}
\usepackage{times}
\usepackage{doi}
\usepackage{patchcmd}
\hypersetup{
    colorlinks=true,
    linkcolor=black,
    citecolor=black,
    filecolor=black,
    urlcolor=black,
}

\title{A Time Projection Chamber\\
with GEM-Based Readout\\
~ \\
\normalsize The LCTPC Collaboration:\vspace{-3ex}}
\author[2]{David Atti\'e}
\author[4]{Ties Behnke}
\author[1]{Alain Bellerive}
\author[17]{Oleg Bezshyyko}
\author[2,28]{Deb Sankar Bhattacharya}
\author[16,27]{Purba Bhattacharya}
\author[16]{Sudeb Bhattacharya}
\author[4,25]{Stefano Caiazza}
\author[2]{Paul Colas}
\author[7]{Gilles De Lentdecker}
\author[4,29]{Klaus Dehmelt}
\author[20]{Klaus Desch}
\author[4,*]{Ralf Diener}
\author[1,18]{Madhu Dixit}
\author[21]{Ivor Fleck}
\author[8]{Keisuke Fujii}
\author[15]{Takahiro Fusayasu}
\author[2]{Serguei Ganjour}
\author[19]{Yuanning Gao}
\author[15,26]{Philippe Gros}
\author[1]{Peter Hayman}
\author[11]{Vincent Hedberg}
\author[15]{Katsumasa Ikematsu}
\author[11]{Leif J\"onsson}
\author[20]{Jochen Kaminski}
\author[9]{Yukihiro Kato}
\author[5,23]{Shin-ichi Kawada}
\author[20,23]{Martin Killenberg}
\author[4]{Claus Kleinwort}
\author[8]{Makoto Kobayashi}
\author[17]{Vladyslav Krylov}
\author[19]{Bo Li}
\author[19]{Yulan Li}
\author[11]{Bj\"orn Lundberg}
\author[20]{Michael Lupberger}
\author[16]{Nayana Majumdar}
\author[8]{Takeshi Matsuda}
\author[1]{Rashid Mehdiyev}
\author[11]{Ulf Mj\"ornmark}
\author[4]{Felix M\"uller}
\author[4,24]{Astrid M\"unnich}
\author[16]{Supratik Mukhopadhyay}
\author[8]{Tomohisa Ogawa}
\author[11]{Anders Oskarsson}
\author[11]{Lennart \"Osterman}
\author[3]{Daniel Peterson}
\author[2]{Marc Riallot}
\author[4]{Christoph Rosemann}
\author[14]{Stefan Roth}
\author[4]{Peter Schade}
\author[4]{Oliver Sch\"afer}
\author[12]{Ronald Dean Settles}
\author[21]{Amir Noori Shirazi}
\author[11]{Oxana Smirnova}
\author[15]{Akira Sugiyama}
\author[5]{Tohru Takahashi}
\author[8]{Junping Tian}
\author[13]{Jan Timmermans}
\author[2]{Maksym Titov}
\author[4]{Dimitra Tsionou}
\author[4,22]{Annika Vauth}
\author[2]{Wenxin Wang}
\author[10]{Takashi Watanabe}
\author[21]{Ulrich Werthenbach}
\author[7]{Yifan Yang}
\author[19]{Zhenwei Yang}
\author[7]{Ryo Yonamine}
\author[4]{Klaus Zenker}
\author[6]{Fan Zhang}

\affil[*] {corresponding author, ralf.diener@desy.de\hfill~}
\affil[1] {Carleton University, Department of Physics, 1125 Colonel By Drive, Ottawa, ON, K1S 5B6, Canada\hfill~}
\affil[2] {CEA Saclay, IRFU, F-91191 Gif-sur-Yvette, France\hfill~}
\affil[3] {Cornell University, Laboratory for Elementary-Particle Physics (LEPP), Ithaca, NY 14853, USA\hfill~}
\affil[4] {Deutsches Elektronen-Synchrotron DESY, A Research Centre of the Helmholtz Association, Notkestrasse 85, \hfill~ \authorcr~ 22607 Hamburg, Germany (Hamburg site)\hfill~}
\affil[5] {Hiroshima University, Advanced Sciences of Matter, 1-3-1 Kagamiyama, Higashi-Hiroshima,\hfill~ \authorcr~ Hiroshima 739-8530, Japan\hfill~}
\affil[6] {Hubei University of Technology (HBUT), No.28, Nanli Road, Hong-shan District, Wuchang, Wuhan, Hubei \hfill~ \authorcr~ Province, P. R. China, Postcode: 430068\hfill~}
\affil[7] {Inter University ULB-VUB, Av. Fr. Roosevelt 50, B1050 Bruxelles, Belgium\hfill~}
\affil[8] {High Energy Accelerator Research Organization, KEK, 1-1 Oho, Tsukuba, Ibaraki 305-0801, Japan\hfill~}
\affil[9] {Kinki University, Department of Physics, 3-4-1 Kowakae, Higashi-Osaka, Osaka 577-8502, Japan\hfill~}
\affil[10] {Kogakuin University, Division of Liberal Arts, 1-24-2 Nishi-Shinjuku, Shinjuku-ku, Tokyo 163-8677, Japan\hfill~}
\affil[11]{Lunds Universitet, Fysiska Institutionen, Avdelningen f{\"o}r Experimentell H{\"o}genergifysik, Box 118, \hfill~ \authorcr~~ 221 00 Lund, Sweden\hfill~}
\affil[12]{Max-Planck-Institut f{\"u}r Physik (Werner-Heisenberg-Institut), F{\"o}hringer Ring 6, 80805 M{\"u}nchen, \hfill~ \authorcr~~ Germany\hfill~}
\affil[13]{Nikhef, National Institute for Subatomic Physics, P.O. Box 41882, 1009 DB Amsterdam, Netherlands\hfill~}
\affil[14]{Rheinisch-Westf{\"a}lische Technische Hochschule (RWTH), Physikalisches Institut, Physikzentrum, \hfill~ \authorcr~~ Otto-Blumenthal-Stra{\ss}e, 52056 Aachen\hfill~}
\affil[15]{Saga University, Department of Physics, 1 Honjo-machi, Saga-shi, Saga 840-8502, Japan\hfill~}
\affil[16]{Saha Institute of Nuclear Physics, 1/AF, Sector 1, Bidhan Nagar, Kolkata 700064, India\hfill~}
\affil[17]{Taras Shevchenko National University of Kyiv, 64/13, Volodymyrska Street, City of Kyiv, Ukraine, 01601\hfill~}
\affil[18]{TRIUMF, Vancouver, BC, V6T 2A3, Canada \hfill~}
\affil[19]{Tsinghua University, Center for High Energy Physics (TUHEP), Beijing, China 100084\hfill~}
\affil[20]{Universit{\"a}t Bonn, Physikalisches Institut, Nu{\ss}allee 12, 53115 Bonn, Germany\hfill~}
\affil[21]{Universit{\"a}t Siegen, Naturwissenschaftlich-Technische Fakult{\"a}t, Department Physik, Emmy Noether \hfill~ \authorcr~~ Campus, Walter-Flex-Str.3, 57068 Siegen, Germany\hfill~}
\affil[22]{\textit{now at} CERN, CH-1211 Gen\`eve 23, Switzerland\hfill~}
\affil[23]{\textit{now at} Deutsches Elektronen-Synchrotron DESY, A Research Centre of the Helmholtz Association, \hfill~ \authorcr~~ Notkestrasse 85, 22607 Hamburg, Germany (Hamburg site)\hfill~}
\affil[24]{\textit{now at} European XFEL GmbH, Notkestrasse 85, 22607 Hamburg, Deutschland\hfill~}
\affil[25]{\textit{now at} Johannes Gutenberg Universit{\"a}t Mainz, Institut f{\"u}r Physik, 55099 Mainz, Germany\hfill~}
\affil[26]{\textit{now at} Laboratoire Leprince-Ringuet (LLR), \'Ecole polytechnique -- CNRS/IN2P3, Route de Saclay,\hfill~ \authorcr~~ F-91128 Palaiseau Cedex, France\hfill~}
\affil[27]{\textit{now at} National Institute of Science Education and Research (NISER) Bhubaneswar, P.O. Jatni, \hfill~ \authorcr~~ Khurda 752050, Odisha, India\hfill~}
\affil[28]{\textit{now at} Saha Institute of Nuclear Physics, 1/AF, Sector 1, Bidhan Nagar, Kolkata 700064, India\hfill~}
\affil[29]{\textit{now at} State University of New York at Stony Brook, Department of Physics and Astronomy, \hfill~ \authorcr~~ Stony Brook, NY 11794-3800, USA\hfill~}

\begin{document}
\maketitle
\begin{abstract}\noindent
For the International Large Detector concept at the planned International Linear Collider, the use of time projection chambers (TPC) with micro-pattern gas detector readout as the main tracking detector is investigated. 
In this paper, results from a prototype TPC, placed in a \SI{1}{\tesla} solenoidal field and read out with three independent GEM-based readout modules, are reported. The TPC was exposed to a \SI{6}{\GeV} electron beam at the DESY II synchrotron. The efficiency for reconstructing hits, the measurement of the drift velocity, the space point resolution and the control of field inhomogeneities are presented. 

\end{abstract}

\section{Introduction}
Time projection chambers (TPC) with micro-pattern gas detector (MPGD) readout are under study for a number of projects in particle and nuclear physics. 
One such project is the International Large Detector (ILD), a detector concept for the planned International Linear Collider (ILC). A TPC is foreseen as ILD's main tracking detector, operated in a magnetic field of \SI{3.5}{\tesla}. The combination of a large instrumented volume, delivering many three-dimensional space points, with a single point resolution of the order of \SI{100}{\um} makes this an attractive and very powerful option. 

A traditional TPC readout with wires would not be able to easily reach this level of resolution. The main reason for this is that the spacing between the wires in the readout module is of order of a millimetre, which in the vicinity of the wire causes the electric and magnetic fields not to be parallel for a similar distance. Because of $\vec{E} \times \vec{B}$ effects, distortions are introduced in the drift paths of the electrons. This spreads the signal electrons along the anode wire introducing an angle which limits the achievable resolution~\cite{LCNote:MWPC}. MPGDs circumvent this problem since the typical length scale of the amplification structure is of the same order as the anticipated resolution, thus reducing the $\vec{E} \times \vec{B}$ effects significantly. 

The LCTPC collaboration~\cite{LCTPC} is investigating the design of such a TPC. Within the collaboration two technical solutions for the gas amplification are being pursued: gas electron multipliers (GEM)~\cite{GEM} and Micromegas~\cite{MICROMEGAS}. They are combined with either a traditional pad-based readout, or with the direct readout by the Timepix chip~\cite{Llopart:2007}. In this paper, results are presented from a study of a prototype time projection chamber equipped with a GEM-based readout combined with a pad plane with pads of pitch size \SI{1.26 x 5.85}{\mm}. 

The performance requirements for the TPC are determined by the requirements coming from the scientific program at the ILC~\cite{ILCTDR}. The detailed study of the properties of the Higgs boson, for example, requires the precise determination of the momentum of charged particles. The TPC alone has to provide a momentum resolution of $\Delta\left(1/p_{\mathrm{T}}\right) = \SI{e-4}{\per\GeV}$. This translates into a single point resolution of $\sim$\SI{100}{\um} over the full drift length of \SI{2.35}{\m}. With $\sim$~200 position measurements along a particle track the TPC offers excellent pattern recognition capability and a tracking efficiency close to \SI{100}{\percent} down to low momenta. In addition, a TPC is capable of providing particle identification information via the measurement of the specific energy loss $\mathrm{d}E/\mathrm{d}x$.

Based on these requirements, a TPC using GEM foils for gas amplification has been developed. In this paper measurements taken with a large prototype are reported, where a large area is instrumented with several readout modules, and the geometry was chosen to be close to the final system planned for the ILD TPC. The prototype chamber has been exposed to an electron test beam in March 2013. Fundamental parameters like point resolution and drift velocity have been measured. Particular emphasis of this study is placed on the determination and treatment of the boundaries between modules, corrections for field distortions, and the demonstration of an alignment procedure based on data for this geometry. The measurements shown in this paper expand earlier work reported in~\cite{DESYGEM}.

In the following, the prototype and the test beam facility at DESY are introduced,  the reconstruction methods used are described 
and the results of the test beam campaign are reported.

\section{Prototype Time Projection Chamber}
The study was done with a prototype TPC, which has been built as part of the EUDET~\cite{EUDET} and the AIDA~\cite{AIDA} project as a shared infrastructure. 

\subsection{The Field Cage}
\label{sec:fieldcage}
The TPC consists of a field cage, a cathode and an anode, containing the readout. The field cage has a cylindrical shape of \SI{77}{\cm} outer diameter and a total length of \SI{61}{\cm}. The maximum drift length in the sensitive volume is nearly \SI{57}{\cm}. The cathode is unsegmented and can provide over \SI{20}{kV} of cathode potential. The field cage contains concentric electrode strips on the inside, which grade the potential from the cathode to the anode. The strips have a width of \SI{2.3}{\mm} and are separated by \SI{0.5}{\mm} wide gaps. The potential of each strip is defined by a cascade of surface mount resistors, soldered to the field shaping electrodes. Separated by a \SI{50}{\um} thick Kapton\textsuperscript{\textregistered} foil, an identical set of electrodes, shifted by half a period, provide mirror strips and ensure that the distortions introduced from the ground potential present on the outside of the field cage are minimised (for more details see~\cite{Schade}). 

\begin{figure}[t]
\begin{subfigure}[b]{0.3945\textwidth}
\iftoggle{blackandwhite}{\includegraphics[width=\textwidth]{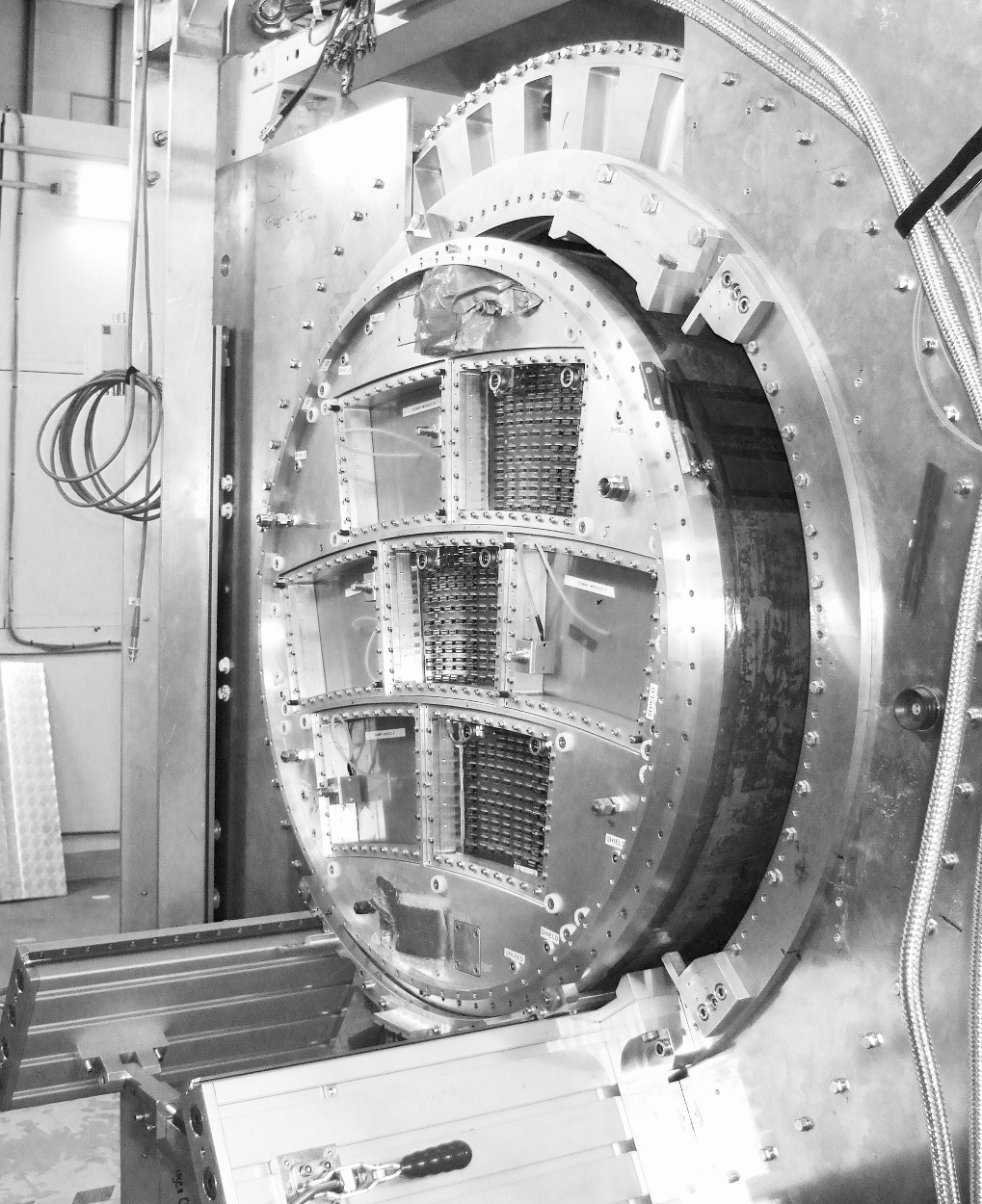}}{\includegraphics[width=\textwidth]{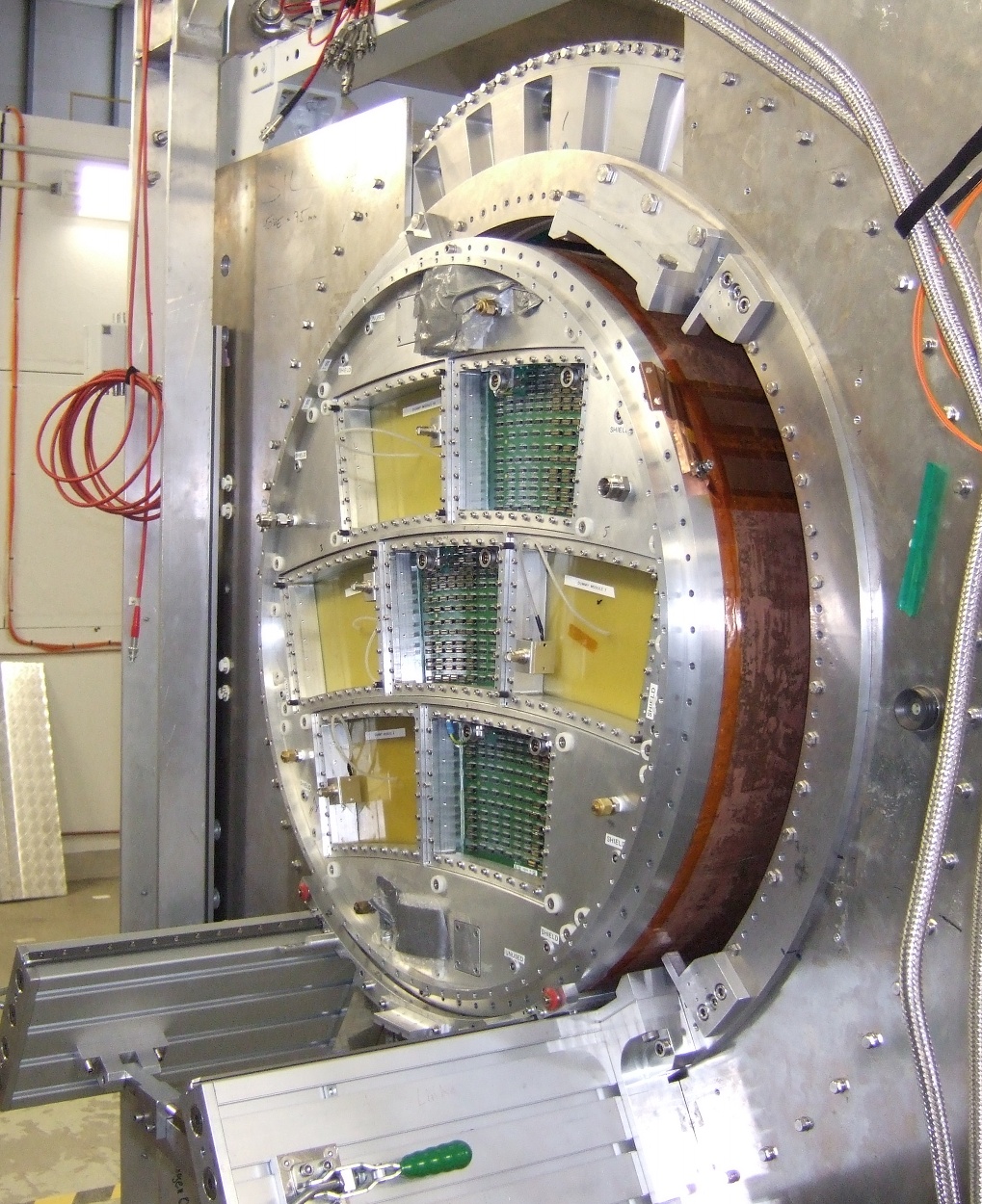}}
\caption{}
\label{sfig:fieldcage}
\end{subfigure}
\hfill
\begin{subfigure}[b]{0.48\textwidth}
\iftoggle{blackandwhite}{\includegraphics[width=\textwidth]{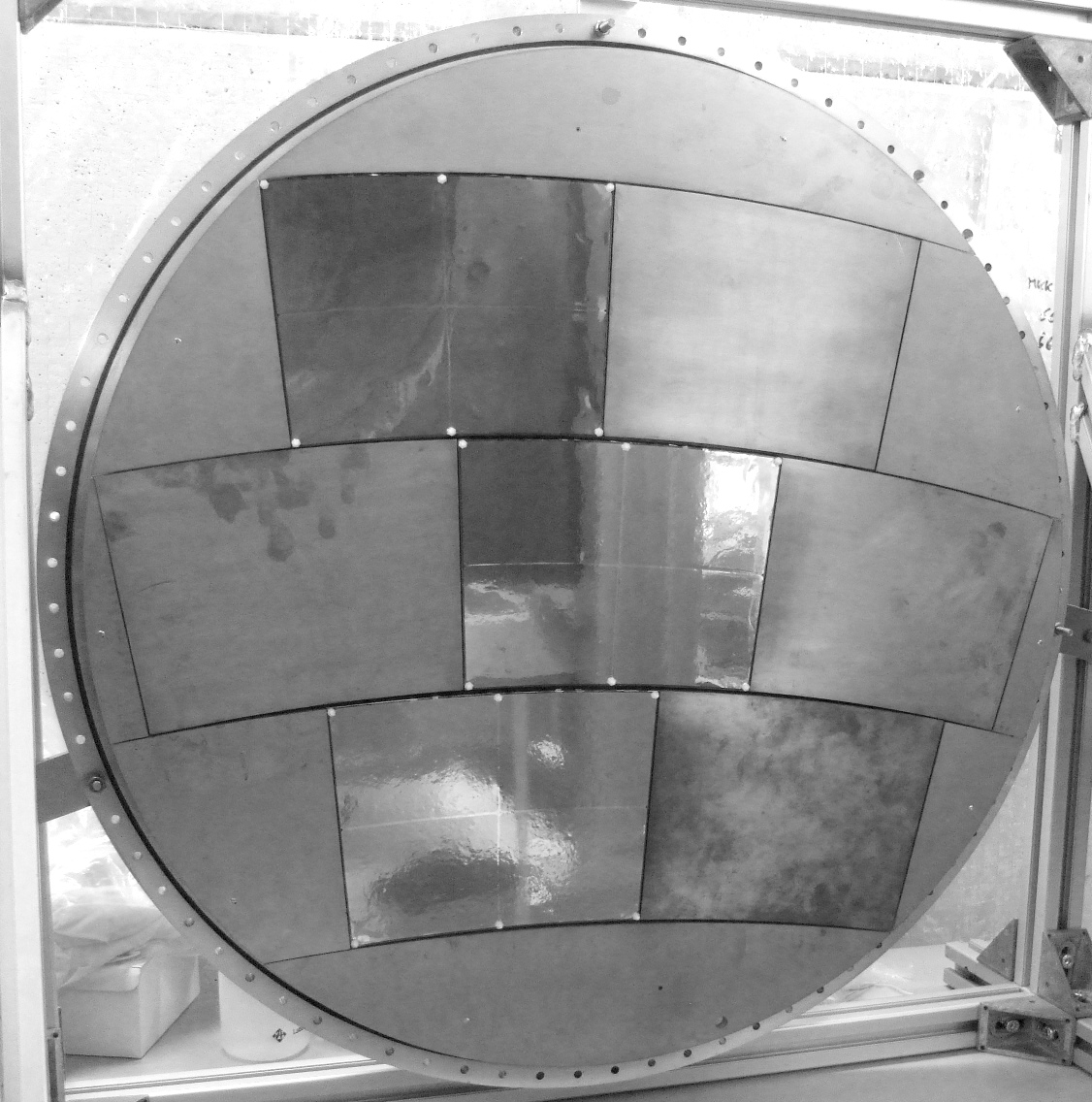}}{\includegraphics[width=\textwidth]{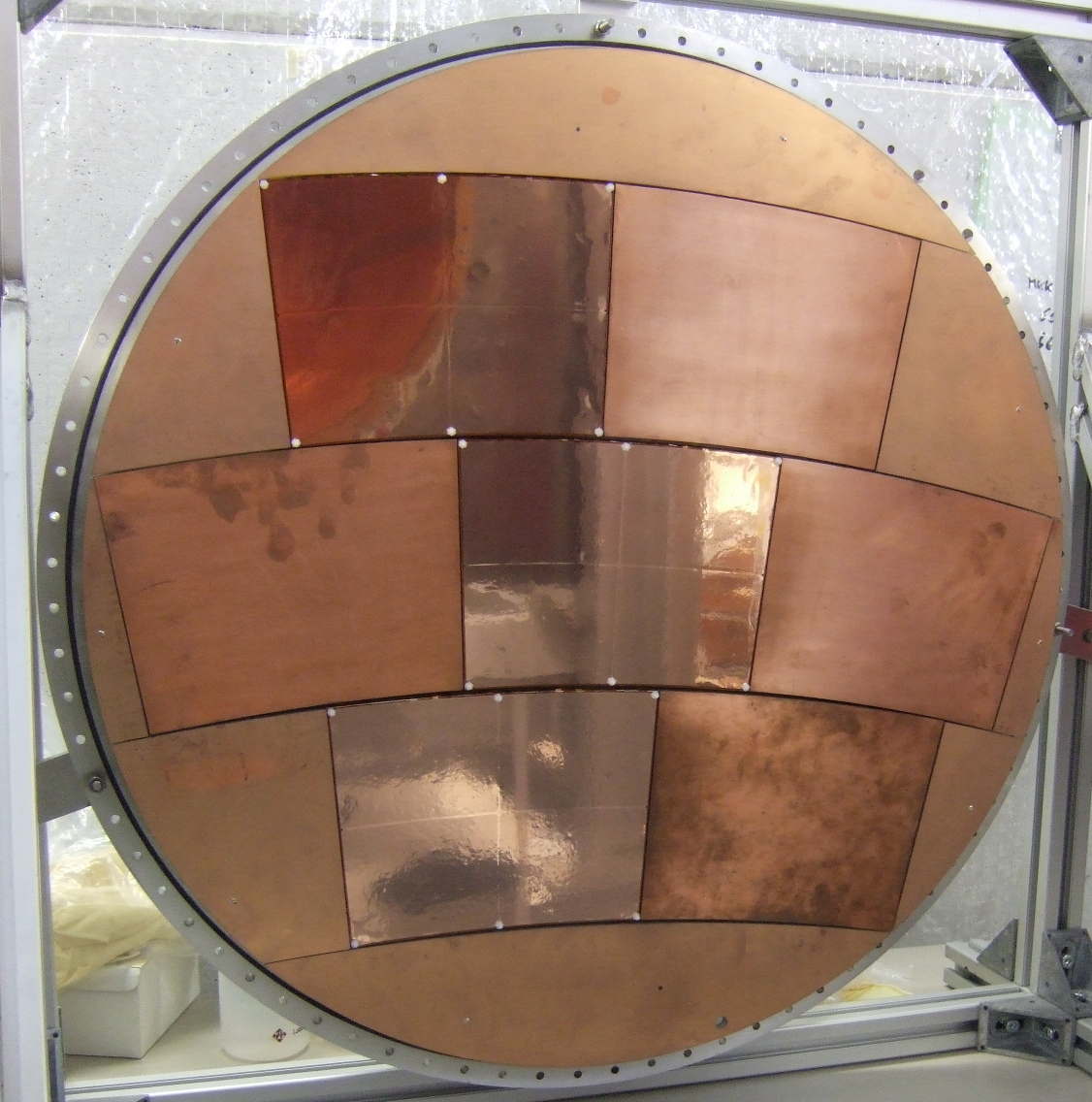}}
\caption{}
\label{sfig:endplateShiny}
\end{subfigure}
\caption{\small \protect\subref{sfig:fieldcage})~View of the field cage inserted into the magnet and equipped with the 7-module endplate.
\protect\subref{sfig:endplateShiny})~Three GEM modules were installed in the endplate, visible in the picture as the shiny surfaces (seen from the inside of the TPC). The other openings are filled with so-called dummy modules.}
\label{fig:endplate}
\end{figure}

An anode endplate made from aluminium was developed at Cornell University. It can support up to seven geometrically identical readout modules, arranged in three rows (see figure~\ref{sfig:fieldcage} and~\ref{sfig:endplateShiny}). All rows have the same radius of about \SI{1.5}{\m} so that all modules have the same shape and are interchangeable. Each module is slightly wedge-shaped and has a size of approximately \SI{23 x 17}{\cm}.  Precise alignment between the modules and the endplate is provided by a set of two precision alignment pins. The modules are inserted from the inside of the field cage, and are pulled against the endplate, where gas tightness is ensured by an O-ring. The spaces between the modules and the wall of the field cage are filled with copper electrodes, which have the same height as the GEM modules, to provide a flat and electrically uniform surface towards the inside of the TPC.

\subsection{The GEM Module}
\begin{figure}[tb]
\begin{subfigure}[b]{0.47\textwidth}
\iftoggle{blackandwhite}{\includegraphics[width=\textwidth]{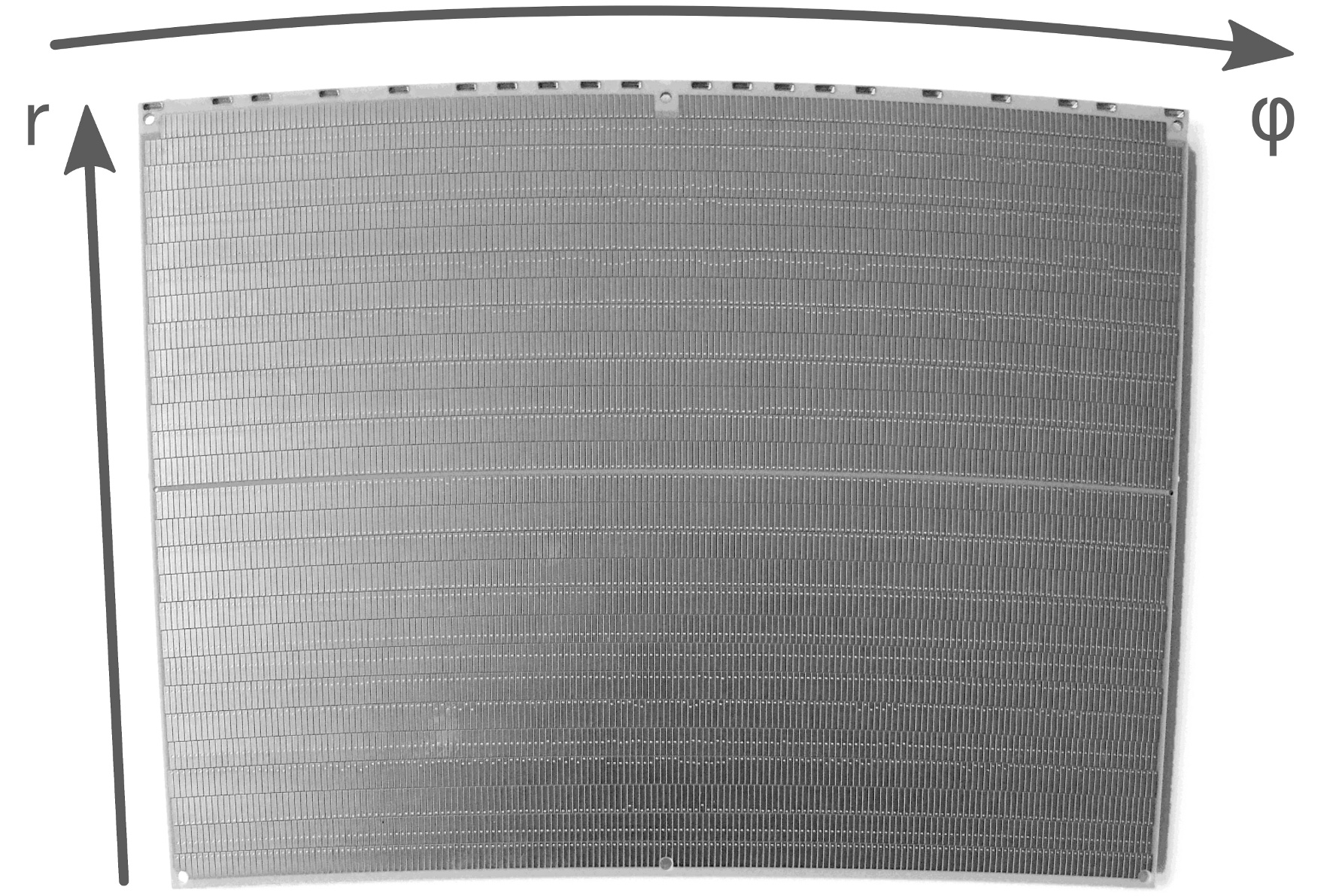}}{\includegraphics[width=\textwidth]{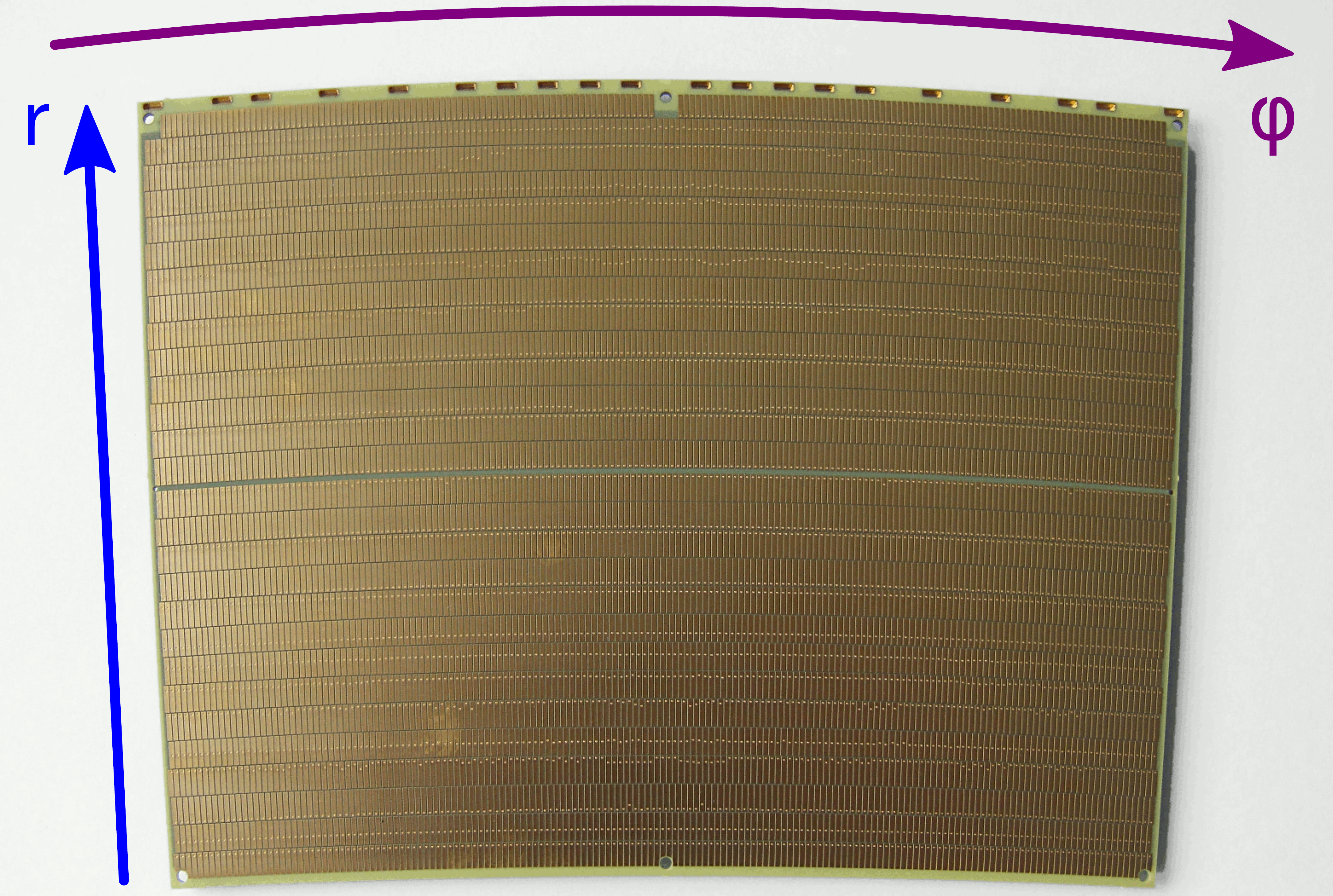}}
\caption{}
\label{sfig:modulePads}
\end{subfigure}
\hfill
\begin{subfigure}[b]{0.47\textwidth}
\iftoggle{blackandwhite}{\includegraphics[width=\textwidth]{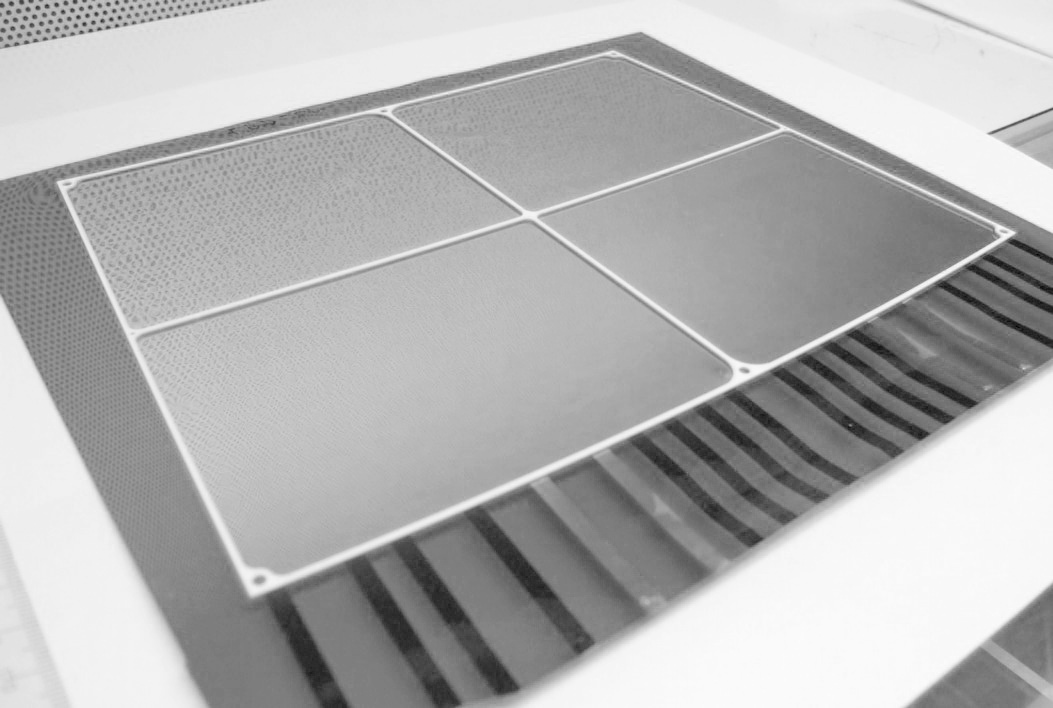}}{\includegraphics[width=\textwidth]{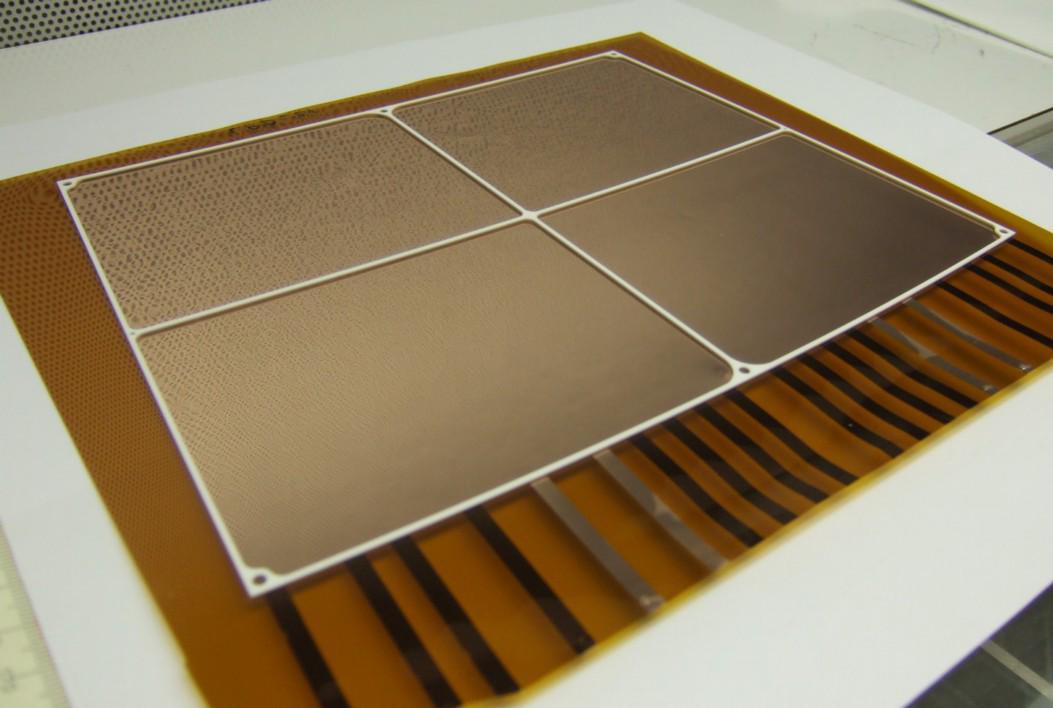}}
\caption{}
\label{sfig:moduleGEM}
\end{subfigure}
\caption [Readout Module GEM]{\small \protect\subref{sfig:modulePads})~Readout pad plane including directions of coordinates and \protect\subref{sfig:moduleGEM})~GEM foil with ceramic frame support used in the construction of the modules.}
\label{fig:moduleParts}
\end{figure}

The modules themselves are built around an aluminium frame, which is responsible for the overall mechanical stiffness. This frame houses the O-ring and the alignment pins. A readout pad plane (figure~\ref{sfig:modulePads}) is glued to the aluminium frame. 
The pad plane was designed in such a way that the inside of the module is covered to nearly \SI{100}{\percent} with pads. The pad plane is realised as a multi-layer printed circuit board. Electrical lines connect each pad to a set of 152 miniature 40-pin connectors~\cite{Connector} on the back side of the pad plane, to which the readout electronics is connected. For the measurements reported in this paper, pads at a pitch of \SI{1.26 x 5.85}{\mm} were used, with an actual pad size of $ \SI{1.06}{\mm} \times \SI{5.65}{\mm}$.
A full pad plane in this granularity has 4828 wedge-shaped pads arranged in 28 circle segments (rows), which share the same origin.

The amplification system consists of a triple GEM stack. The GEMs are custom-tailored to have the same wedge shape as the modules, as shown in figure~\ref{sfig:moduleGEM}. They are based on the well proven CERN GEM design with \SI{50}{\um} insulator thickness. The distance between adjacent holes is \SI{140}{\um}, the hole diameter is \SI{70}{\um}. The double conical holes are arranged in a hexagonal pattern. To limit the charge transfer in case of a discharge in a GEM, the side of the GEM facing the pad plane is segmented into four sectors. 
The side of the GEMs facing inside the TPC volume is unsegmented to minimise electric field distortions in the sensitive volume. Electrically, all sides and sectors of each GEM are powered individually and are protected by \SI{10}{\Mohm} resistors.

\begin{figure}[tb]
\begin{subfigure}[b]{0.47\textwidth}
\iftoggle{blackandwhite}{\includegraphics[width=\textwidth]{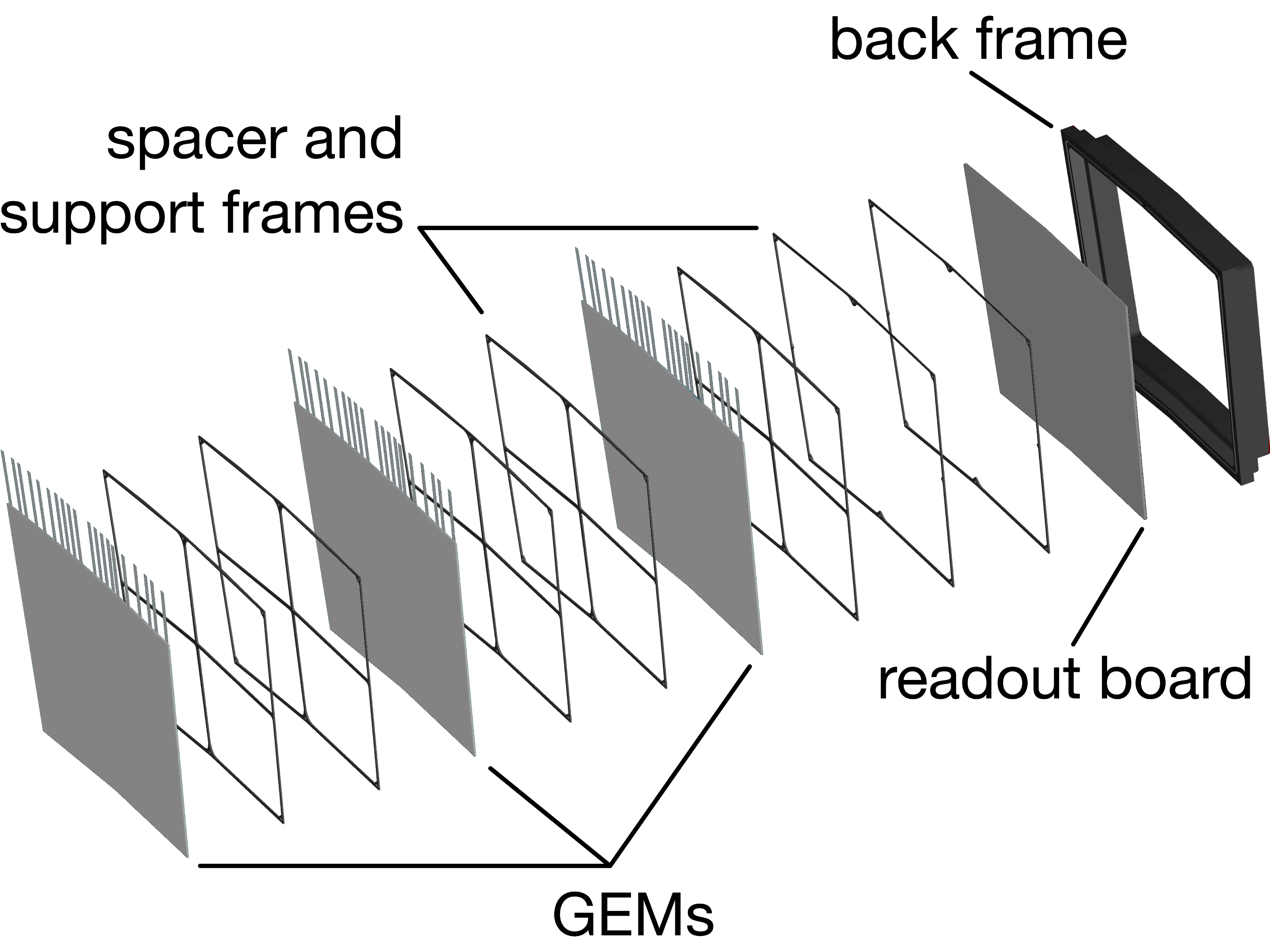}}{\includegraphics[width=\textwidth]{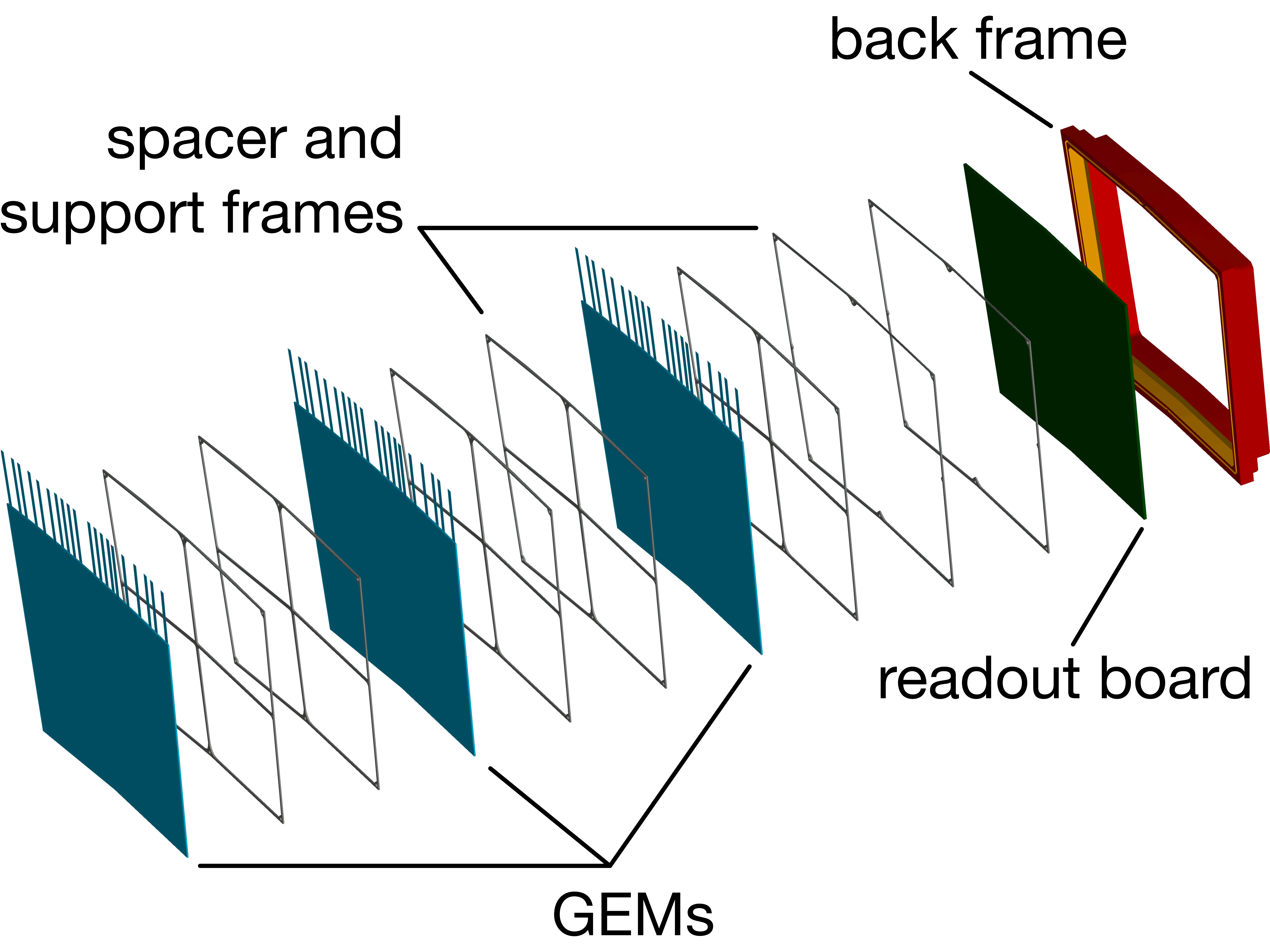}}
\caption{}
\label{sfig:moduleExp}
\end{subfigure}
\hfill
\begin{subfigure}[b]{0.47\textwidth}
\iftoggle{blackandwhite}{\includegraphics[width=\textwidth]{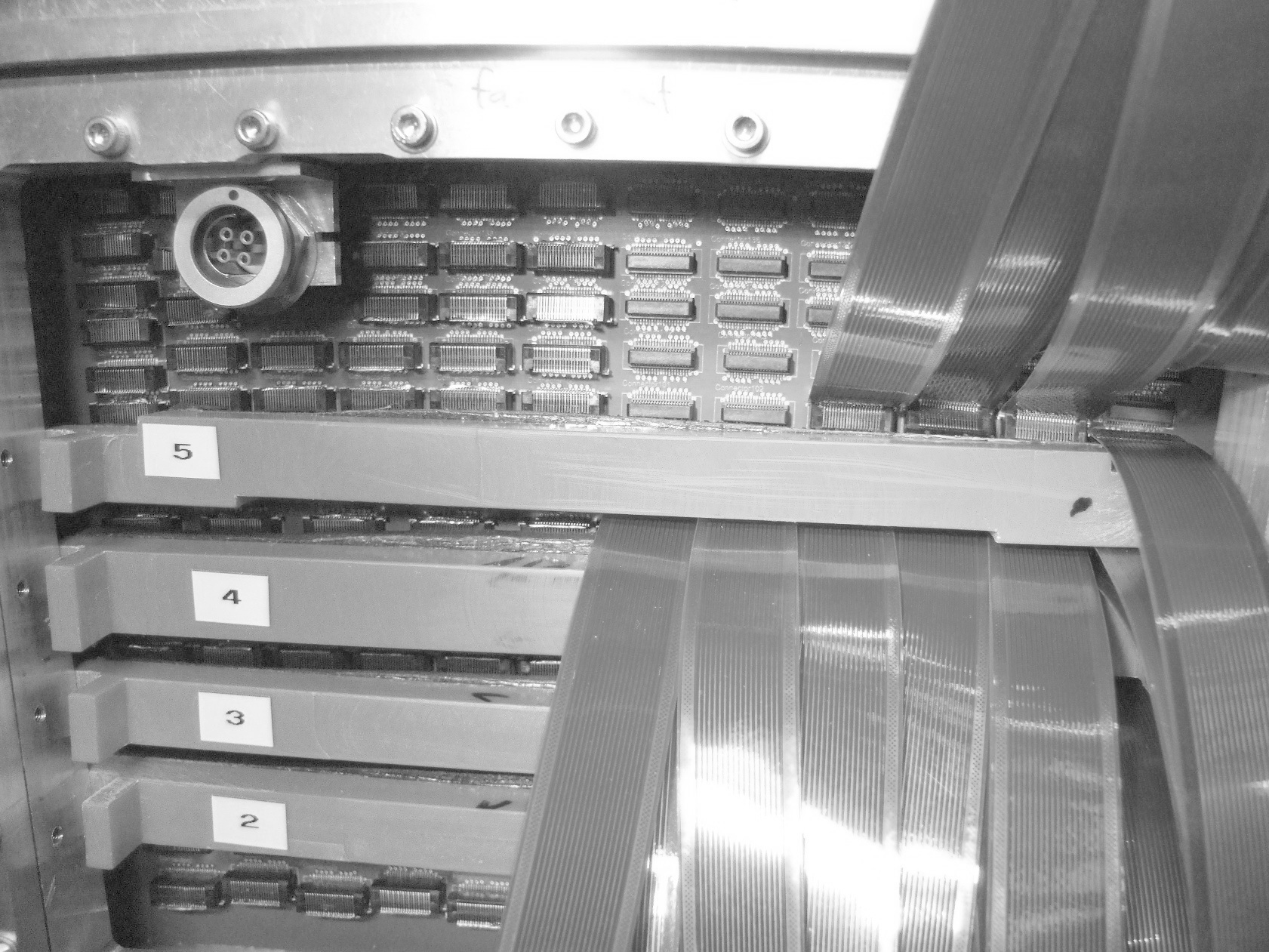}}{\includegraphics[width=\textwidth]{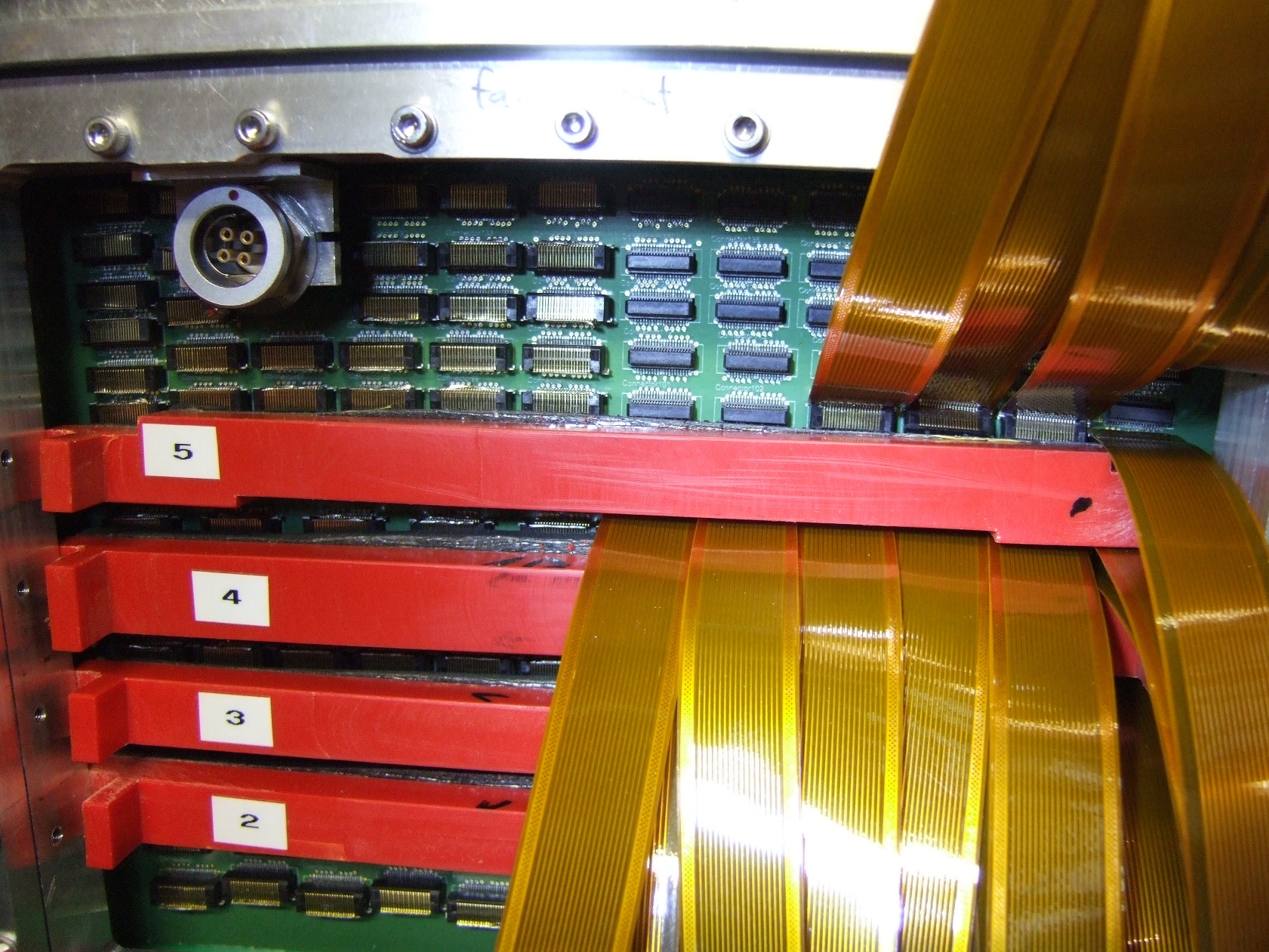}}
\caption{}
\label{sfig:moduleBack}
\end{subfigure}
\caption [Readout Module GEM]{\small \protect\subref{sfig:moduleExp})~Exploded view of one module showing the sequence of GEM foils and ceramic frames. \protect\subref{sfig:moduleBack})~Partly connected module with one of the high voltage connectors visible on the top left and some of the Kapton cables for the pad signal readout on the right. }
\label{fig:moduleAssembled}
\end{figure}

The GEMs are glued to a ceramic frame. This frame provides mechanical stiffness to the GEM and, at the same time, acts as a spacer between the GEMs in a stack. The frames have a thickness of \SI{1.4}{\mm}, and the individual partitions have a width of \SI{1}{\mm}. The partitions coincide with the electrical separations on the GEM. 
By stacking up several of these frames, as illustrated in figure~\ref{sfig:moduleExp}, different spacings between the GEMs can easily be realised. The frames are sufficiently stable and provide enough support to ensure a flat GEM surface, without applying large mechanical tension (for details see~\cite{1748-0221-8-12-P12009,HallermannPhD}). The glueing of the frames to the GEMs was done in a semi-automatic setup with a glueing robot, which dispensed glue in a carefully metered way. This is a particularly important step as the integrity of the glue joint between the GEM and the ceramic frame, and between the readout pad plane and the ceramic frame, is an integral part of the high voltage system. 
Failures of this glue joint can lead to discharges from the GEM surfaces to ground or to the high voltage connections, which are present outside the modules. In addition, glue spillover into the nearby GEM holes can lead to a reduced high voltage stability in these areas.

All high voltage connections of a GEM are brought to the outer radius of the GEM. They are then connected with flat Kapton cables to the pad plane, through which they are routed to two multi-pin high voltage connectors at the top side of the module, as can be seen in figure~\ref{sfig:moduleBack}.

The complete readout is formed by several modules. In the final position, a gap of a few millimetres width separates the modules on the inside of the TPC. The width of the gap on the top and bottom side varies between roughly \SI{3}{\mm} and \SI{4}{\mm} depending on the exact position. The gap on the right and left side is \SI{1}{\mm} wide. These gaps have a significant effect on the field quality close-by. To control and minimise these effects, a field shaping electrode is installed on the topmost ceramic frame of the GEM stack, running along the left, right and bottom sides of the module.\footnote{On the top side of the module, the high voltage connections of the GEMs are brought down on its side to the readout pad plane. Therefore, the field shaping electrode could not be installed on this side, too. The distortions in this gap are controlled by the field shaping electrode of the neighbouring module.} The potential applied to this electrode can be controlled separately. 
It is optimised to maximise the charge collection efficiency on pads close to the edge of the modules~\cite{ZenkerPhD}. 

Altogether, three identical modules have been constructed and were installed into the prototype. The other four module openings were filled with so-called dummy modules, i.e.\ modules which simply fill the place of a module with a copper electrode.

\subsection{Readout System}
The TPC is read out with a modified ALTRO readout system~\cite{ALTROchip,altro:eudet08}. Each pad is connected to a programmable PCA16 charge-sensitive preamplifier, which allows to set the gain, the shaping time, the decay time and the polarity of the pulse~\cite{EudetPCA16,ShiMaster}. For the data described in the paper, a gain of \SI[per-mode=symbol]{12}{\mV \per \femto\coulomb} and a peaking time of \SI{120}{\ns} were used. The signal is then digitised with the 16-channel ALTRO chip, which offers a resolution of 10~bit and an event buffer of \SI{1}{k} 10-bit words. The sampling rate can be set to 5, 10, 20, and \SI{40}{\MHz}. The system has a footprint of \SI{1}{\square\cm} per channel, which is significantly larger than one pad. Therefore, the front-end boards are installed in a separate support wheel in front of the TPC, in which they are mounted perpendicularly to the endplate to increase the channel density. The signals from the pad plane are brought via Kapton cables 
to the front-end boards, where 
they undergo amplification, shaping, digitisation, pedestal subtraction and zero suppression before being stored in the event buffer. From there they are transferred via an optical link to a computer and saved to disk.

The system is run in a common stop mode. Upon receiving a signal from the beam trigger, after a proper delay, the readout and digitisation cycle is stopped. It is adjusted in such a way that enough samples are stored to cover the complete drift including a safety margin, plus 15 samples before the trigger which corresponds to about \SI{750}{ns}. 
The digitisation is run at a sampling rate of \SI{20}{\MHz}, since at \SI{40}{\MHz} the standard ALTRO chip could not maintain the desired resolution. The next trigger is only accepted once the output memory has been read out and cleared. Due to the test beam spill structure and with typically only a few tracks per event in the TPC, data was taken at about \SI{50}{\Hz} up to \SI{70}{\Hz}.

For the experiment described in this paper, 7212 pads were read out. They were distributed over the three modules in such way that a fully instrumented ``road'' was available across all three modules (see figure~\ref{sfig:endplateNumbered}), including module boundaries and areas where partitions from the ceramic grids shadow the track signal of the beam. 

\section{Experimental Setup}
The prototype has been exposed to an electron beam at the DESY II test beam fa\-ci\-li\-ty~\cite{DESY2TB}. DESY II provides electron beams of up to \SI{6}{\GeV} at a rate of up to several kHz, depending on the chosen energy. One of three existing beam lines is equipped with the ``PCMAG test infrastructure''. This setup consists of a thin-walled superconducting solenoid provided by KEK, which can provide a magnetic field of up to \SI{1}{\tesla}~\cite{pcmag:magnet}. The magnet is mounted on a movable stage, which allows the setup to be moved horizontally and vertically, perpendicular to the beam line, as well as rotate by $\pm \SI{45}{\degree}$ in the horizontal plane. The stage can position the device under test with a precision of about \SI{0.2}{\mm} horizontally, \SI{0.1}{\mm} vertically, and within \SI{0.1}{\degree} in angle. 

The magnet has a bore with a diameter of \SI{85}{\cm}, and a usable magnetic length of \SI{1.1}{\m}. Since the magnet is not equipped with an iron return yoke, the field in particular close to the ends of the magnet is rather inhomogeneous and acquires a large radial component. A precision field map of the magnet was recorded using a movable measurement head in 2008~\cite{pcmag:fieldmeas,pcmag:fieldana}. Continuous measurements of the field strength in a number of locations ensure that overall changes of the magnetic field can be tracked. 

Inside the bore of the magnet, a rail system is installed on which test devices can be mounted at different positions within the magnet. The large TPC prototype is supported on a sled, which can move in and out of the magnet and can be used to rotate the chamber around the magnetic field axis. 

Usually the magnet is positioned perpendicular to the beam. The walls of the magnet present about \SI{20}{\percent} of a radiation length, so that an electron beam of \SI{6}{\GeV} easily penetrates the magnet and the device under test. A set of four consecutive scintillation counters, of which each has an area of approximately \SI{2.5 x 2.5}{\cm}, is mounted about \SI{1.5}{\m} in front of the magnet. The coincidence between them is used as a beam trigger. In addition, a second set of scintillation counters above and below the magnet provide a cosmic trigger for tests without beam.

A slow control system monitors environmental parameters, such as the gas quality and the electric field settings, and is used to deliver information on the state of the magnet. The slow control system uses the DOOCS control software~\cite{TPCslowcontrol}. 

The chamber has been operated for these measurements with a gas mixture of 95\% argon, 3\% tetrafluoromethane (CF$_4$), and 2\% isobutane (iC$_4$H$_{10}$). The gas quality was constantly monitored during the measurement. The gas volume in the chamber was exchanged typically every six hours, the used gas was vented. For the results reported in this paper the oxygen contamination was around \SI{50}{ppm}, and the water content in the gas was around \SI{60}{ppm}. The chamber was operated at atmospheric pressure. Ambient temperature and pressure were constantly monitored. 

\begin{figure}[htb!]
\begin{subfigure}[b]{0.47\textwidth}
\iftoggle{blackandwhite}{\includegraphics[width=\textwidth]{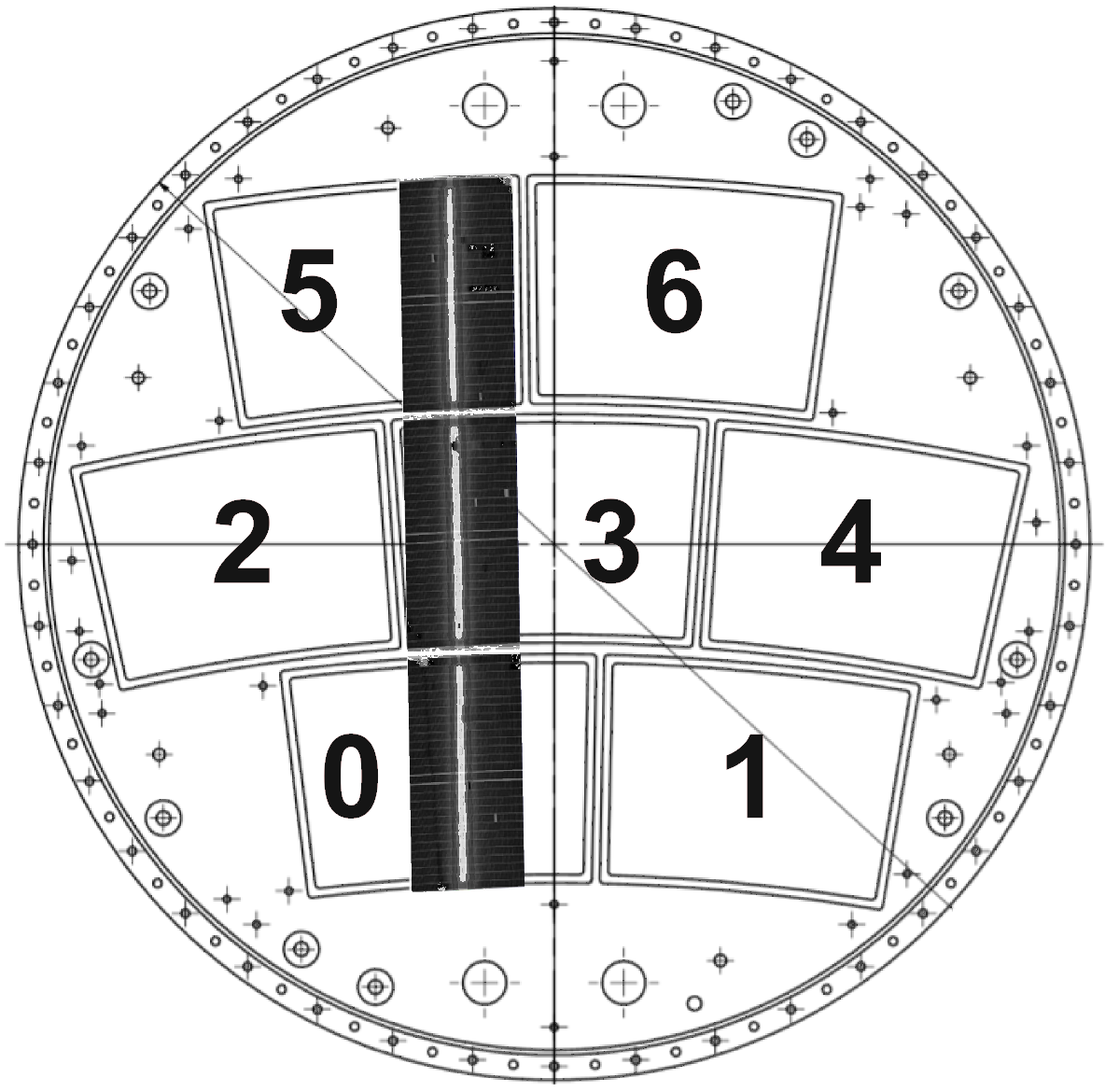}}{\includegraphics[width=\textwidth]{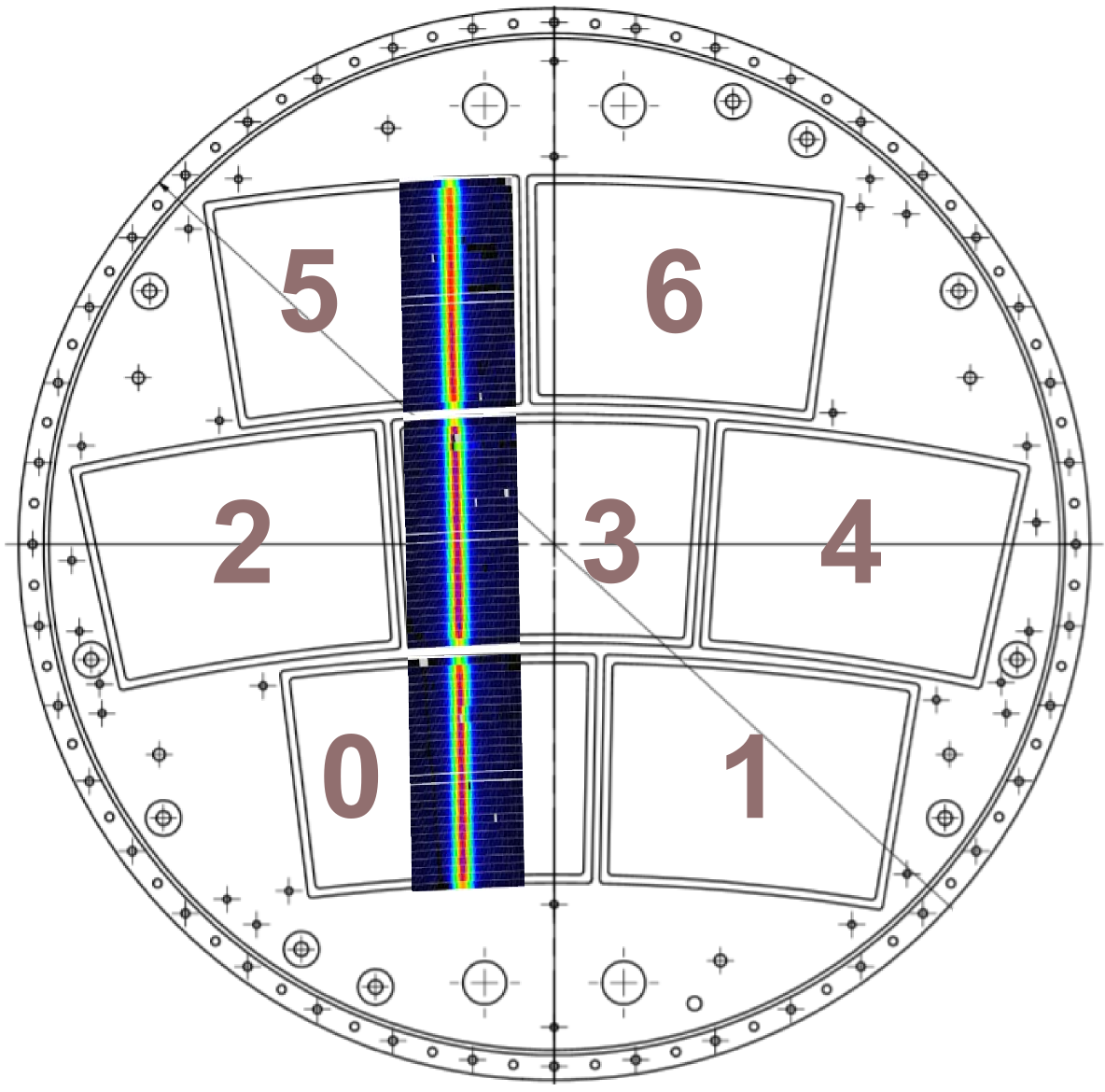}}
\caption{}
\label{sfig:endplateNumbered}
\end{subfigure}
\hfill
\begin{subfigure}[b]{0.47\textwidth}
\iftoggle{blackandwhite}{\includegraphics[width=\textwidth]{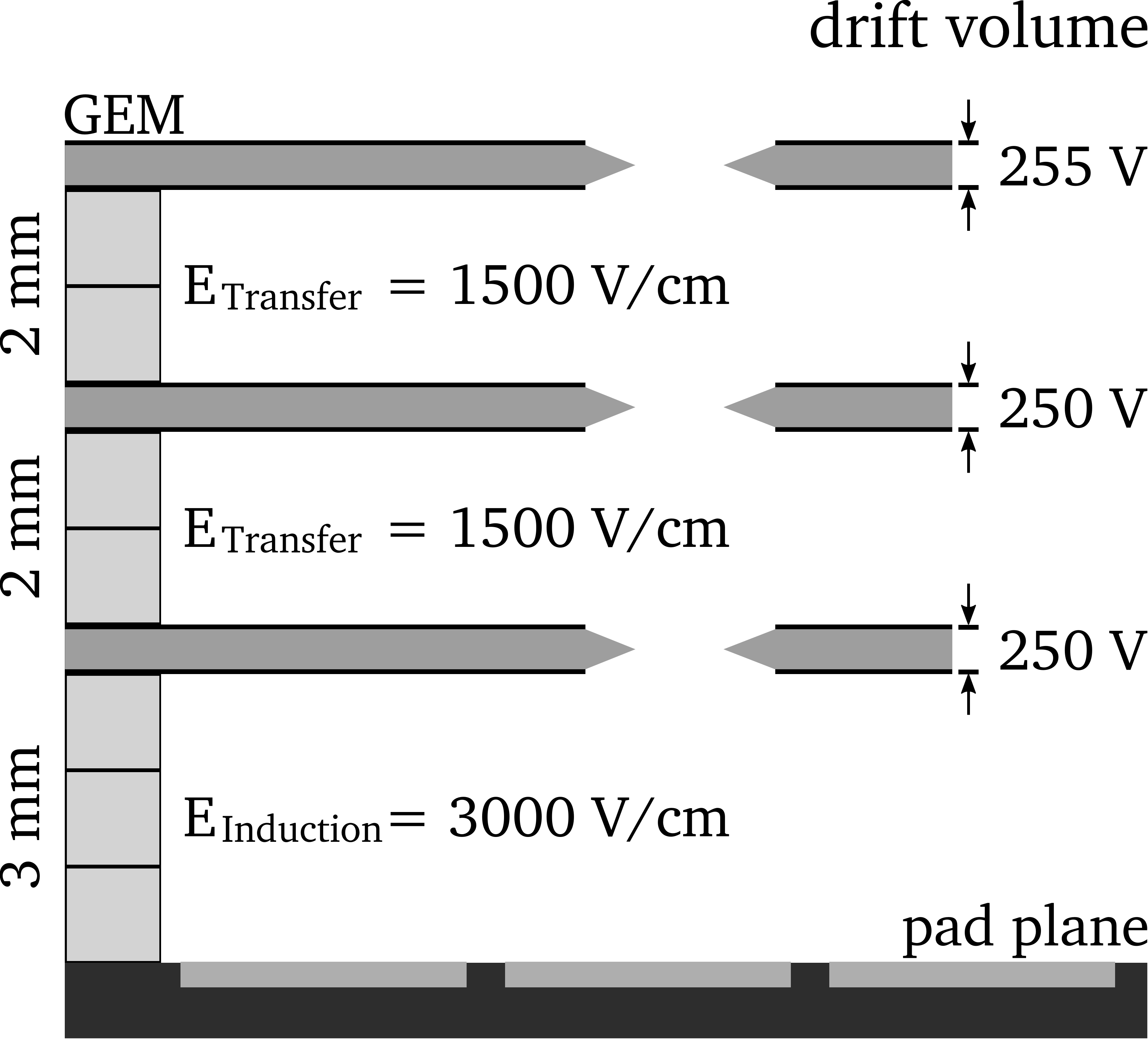}}{\includegraphics[width=\textwidth]{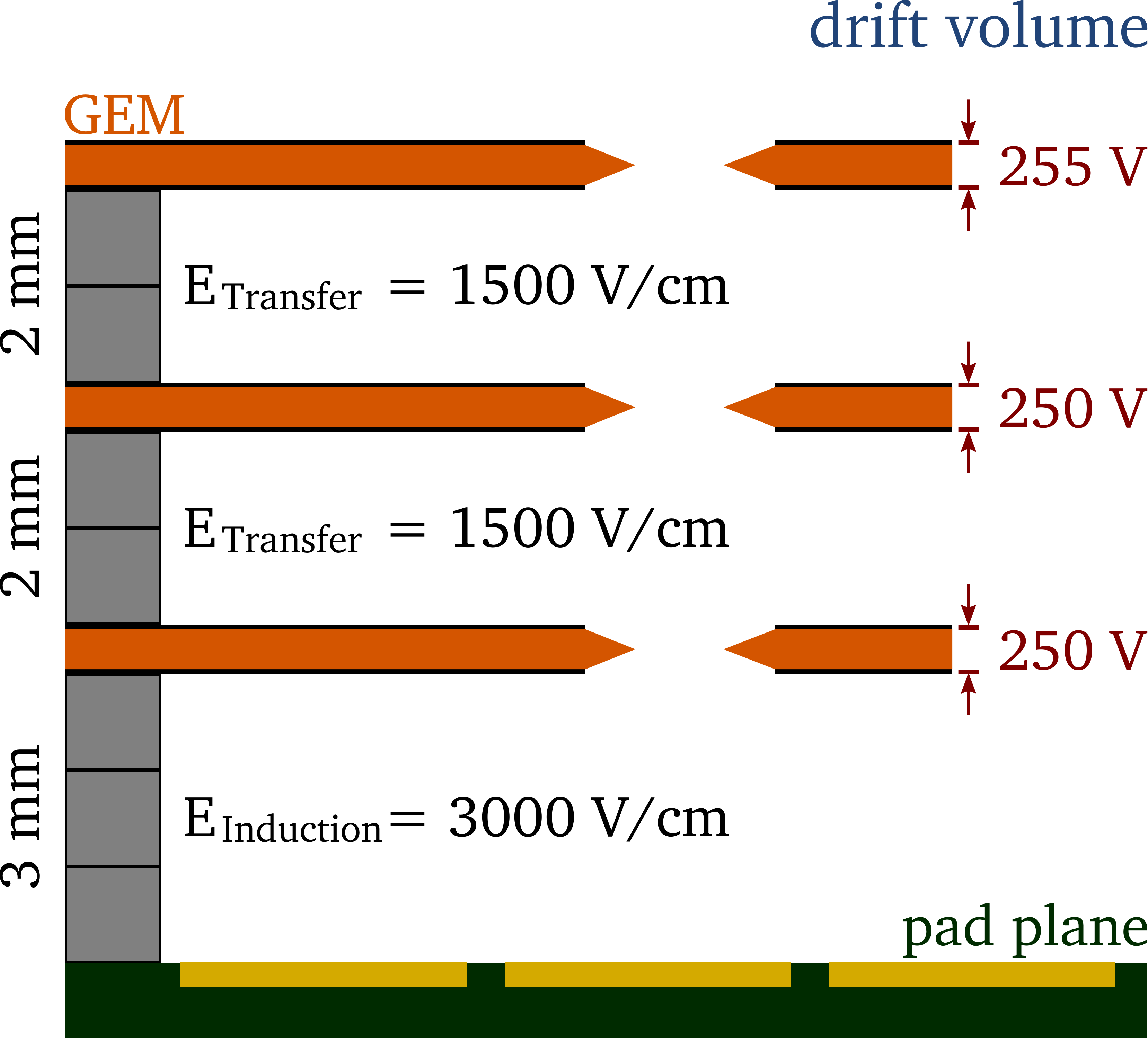}}
\caption{}
\label{sfig:gemVE}
\end{subfigure}
\caption{\small \protect\subref{sfig:endplateNumbered})~Profile of the beam, from an overlay of a full measurement run, superimposed onto a drawing of the endplate (seen from the inside of the TPC). The dark area shows the instrumented area. \protect\subref{sfig:gemVE})~Configuration of the GEM stack. }
\label{fig:TB}
\end{figure}

\sisetup{per-mode=symbol} 
The TPC was usually operated at a drift field of \SI{240}{\V \per \cm}, i.e.\ at the maximum of the drift velocity versus drift field relation for the used gas mixture. A few measurements were done at a reduced drift field of \SI[per-mode=symbol]{130}{\V \per \cm}, i.e.\ at the point of minimal transverse diffusion.
Figure \ref{sfig:gemVE} shows the configuration of the GEM stack: The potential across the two GEMs closest to the pad plane was \SI{250}{\V}, the topmost GEM was operated at \SI{255}{\V}. 
The transfer fields between the GEMs were \SI{1500}{\V \per \cm}, the induction field between the last GEM and the pad plane \SI{3000}{\V \per \cm}.
The transfer gaps were set to \SI{2}{\mm}, the induction gap was \SI{3}{\mm} high. This operating point has been shown to allow stable operation at a gain of approximately 2000 for the complete setup. Dedicated measurements of the gain were done with a small setup using \SI{10 x 10}{\cm} CERN GEMs. The parametrised results were used as input for a simulation of the gain based on electrostatic properties of the setup~\cite{ZenkerPhD}. This simulation has been used to estimate the gain quoted above. 
\sisetup{per-mode=reciprocal} 

\section{Reconstruction Methods}
\label{sec:reco}
The reconstruction and analysis of the measured has been performed with the MarlinTPC~\cite{MarlinTPC} software package, which is based on the linear collider software suite~\cite{lcsoft,Marlin,lcio}. In the following, the different steps and results will be described.

\subsection{Hit Reconstruction}
\label{sec:hitreco}
Electrons created in the drift volume of the TPC drift towards the anode. They pass through the GEM stack experiencing avalanche amplification. At the end of this process a charge cloud drifts from the last GEM towards the pad plane. The width of the cloud depends on the initial electron distribution, the transverse diffusion in the gas and the amplification in the stack. The choice of gas and operating point has been done in a way that on average more than three pads in a row see a signal from the charge cloud. 
These signals created on individual pads are called pulses. A row-based clustering algorithm is run over the pulses. The combination of several pulses in a cluster on the pad row is called a hit. The hits are analysed and their position on the pad plane, the timing, and the total charge are reconstructed and stored. 

For each pad, the charge distribution is measured in dedicated runs without beam. The mean of this charge distribution defines the pedestal used for the zero suppression in the readout electronics. The noise width of the pad is determined from the RMS of the charge distribution and usually has a value of 1~ADC count or less. 
To be selected in the reconstruction, a pulse has to cross a threshold which is set at 5~times this noise width. To get a complete time evolution of the signal, 3~time bins before the threshold crossing are saved as well. The pulse stops if the signal dips below a second threshold, and if at least 5~bins were above the threshold in between the start and the stop bin. Pulses from neighbouring pads within a row are combined into hits if they are within a time window of 10~time bins with respect to the time of the largest pulse. The charge of each pulse is calculated as the sum of the ADC counts in the bins, from the start to the stop bin. The coordinate along a pad row is calculated as the average of the charge-weighted position of all pulses contributing to the hit. The time of the hit is then determined from the largest pulse in the hit. It is derived from the inflection point of the rising edge of the pulse. 

Technically, the inflection point is determined from a Gaussian function fitted to the rising edge of the signal plus the four following time bins. Due to the Gaussian-like rise of the pulse from the shaper, fitting a Gaussian function to the rising edge was found to work stably for the electronics and settings used. The inflection point corresponds to the mean of the Gaussian minus the standard deviation of the distribution. This method has been shown to be stable and precise, even in the presence of noise. The time information of the neighbouring pulses is not used, since it is affected by a number of systematic effects. They are systematically earlier in time than the central pulse, due to induced pulses, and also show a strong dependence of the timing information on the total charge. 
The charge of the hit is finally calculated from the sum of the charge of all pulses contributing to the hit. 

\subsection{Track Reconstruction}
\label{sec:trackreco}
The track finding is based on an iterative Hough transformation~\cite{KleinwortHough}. The track parameters are determined using a ``General Broken Lines Fit''~\cite{KleinwortGBL,GBLwiki} on all hits identified as belonging to a track. The General Broken Lines method is mathematically equivalent to a Kalman filter. It is implemented such, that it allows to directly use the Millepede II~\cite{millepedeNIM,millepedeWiki} toolkit for track-based alignment and calibration. Here, it has been used neglecting the material in between hits. As the track model either a straight line is used, for data taken at \SI{0}{\tesla} magnetic field, or a helix for data taken with magnetic field. The track parameters are defined in~\cite{LC-DET-2006-004}.

\subsection{Data Quality Cuts}
\label{sec:cuts}
Combining all three modules, a maximum of 84 hits can be reconstructed on one track, corresponding to the number of rows passed by a track. Due to technical problems, mostly because of faulty or intermittent connectors, 13 of these 84 rows did not work properly and were excluded from the analysis. To ensure a good reconstruction quality, only tracks with at least 60 hits are taken into account.
In addition, all events that contain more than one reconstructed track have been excluded from the analysis to avoid events with tracks from interactions with the magnet or field cage wall. Unless noted otherwise, no further cuts were applied in the following analyses.
                                                                                                                                                                     
\section{Results}
\subsection{Efficiency}
\begin{figure}[tb]
\begin{subfigure}[b]{0.48\textwidth}
\iftoggle{blackandwhite}{\includegraphics[width=\textwidth]{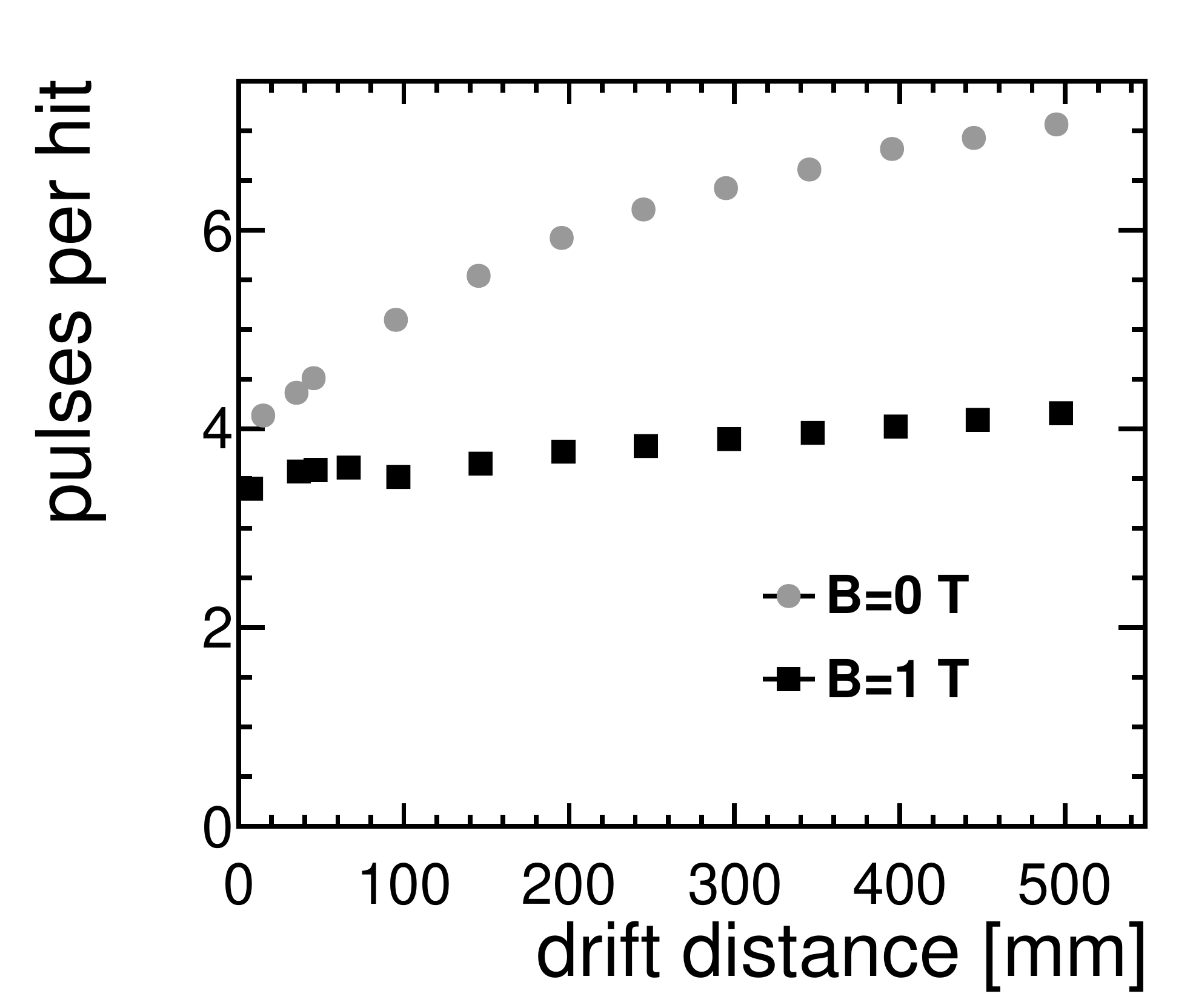}}{\includegraphics[width=\textwidth]{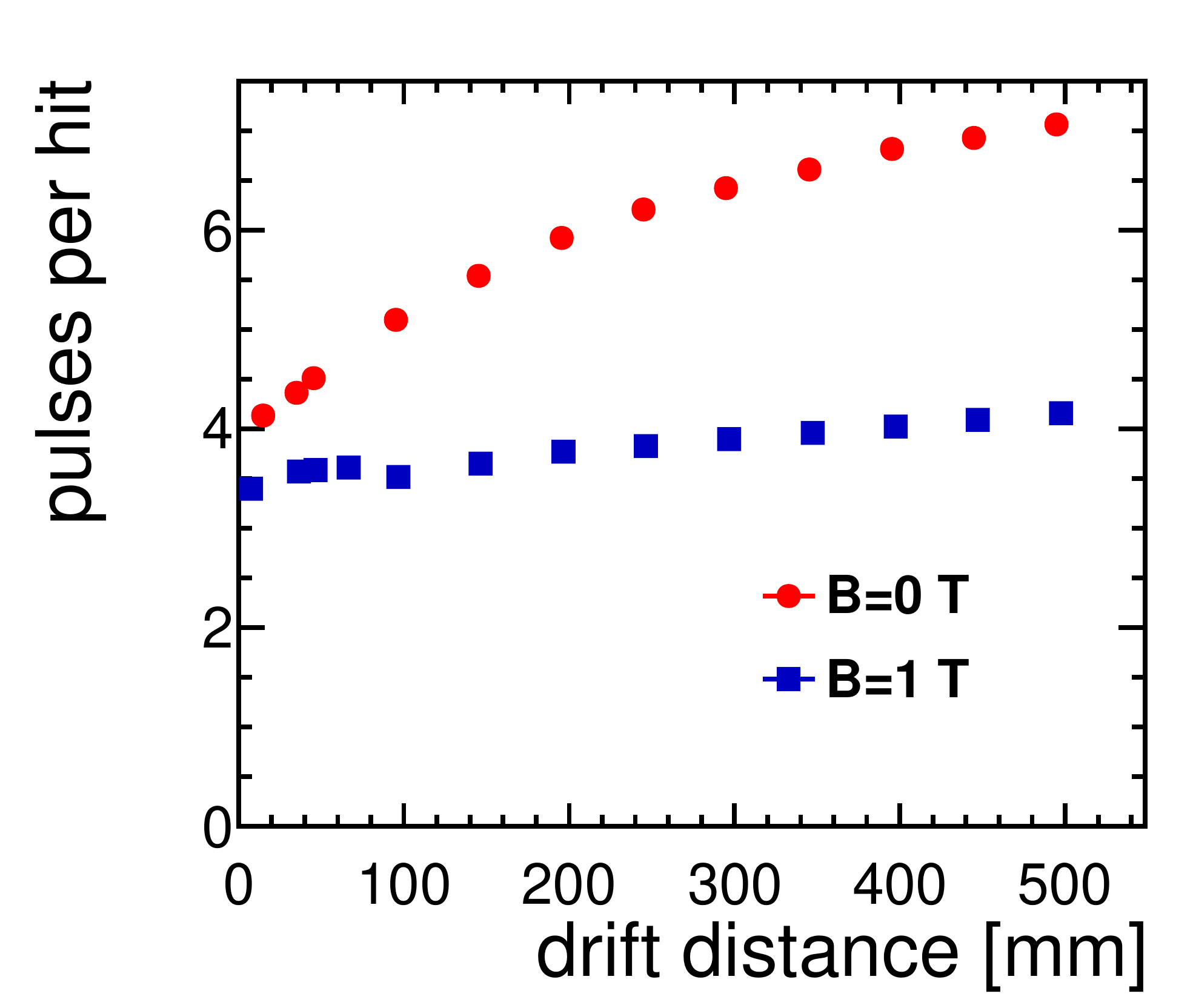}}
\caption{}
\label{sfig:padsphit}
\end{subfigure}
\hfill
\begin{subfigure}[b]{0.48\textwidth}
\iftoggle{blackandwhite}{\includegraphics[width=\textwidth]{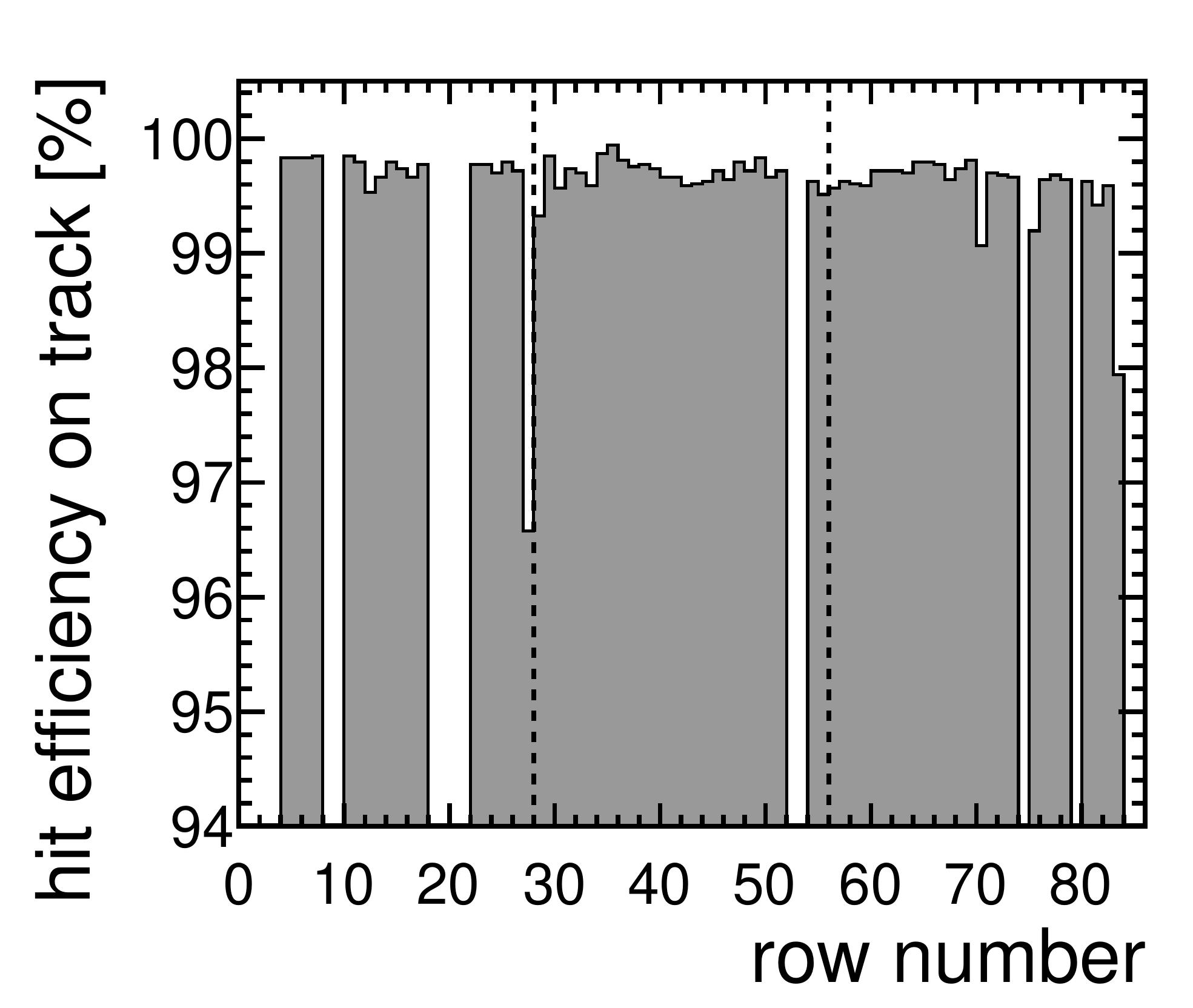}}{\includegraphics[width=\textwidth]{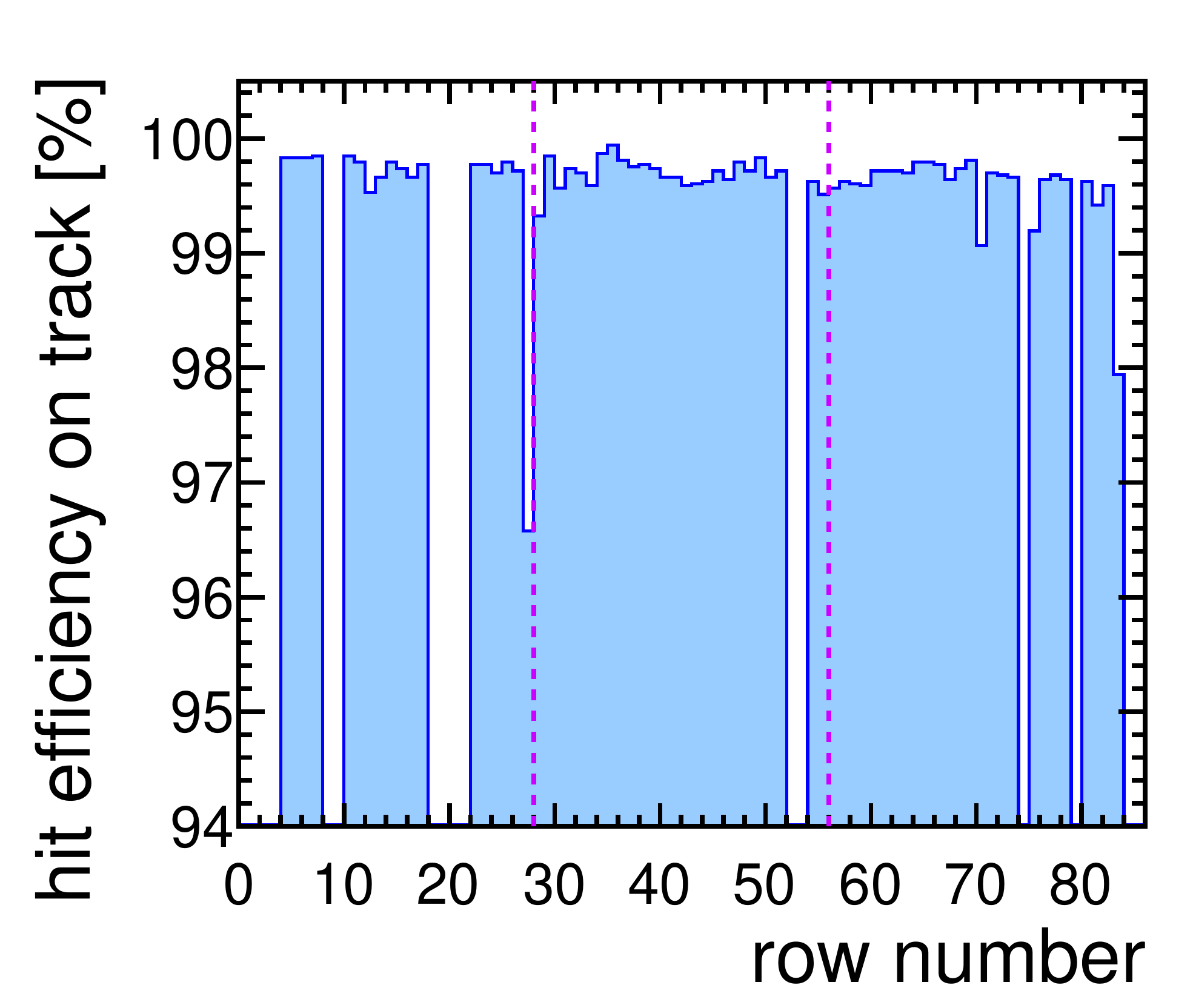}}
\caption{}
\label{sfig:hiteff}
\end{subfigure}
\caption{\small \protect\subref{sfig:padsphit})~Number of pads contributing to a hit versus drift distance, for data taken at \SI{0}{\tesla} and \SI{1}{\tesla} magnetic field. \protect\subref{sfig:hiteff})~Hit efficiency per pad row at \SI{1}{\tesla}. The borders between two modules are marked by dashed lines. The white areas correspond to rows with dead channels in the readout.}
\label{fig:padsHits}
\end{figure}

In figure~\ref{sfig:padsphit} the average number of pads contributing to a hit is shown as a function of the drift distance, without magnetic field and for \SI{1}{\tesla} magnetic field, respectively. The effect of the diffusion is clearly visible since it makes the number of pulses increase with larger drift distance. The effect is significantly reduced once the magnetic field is switched on. 

In figure~\ref{sfig:hiteff}, the hit finding efficiency per pad row is shown for a measurement at \SI{1}{\tesla} magnetic field. It is defined as the ratio between the number of times a pad row participated in a track and the number of tracks which should have produced a hit on that pad row. 
In addition to the data cuts described in section~\ref{sec:cuts}, track candidates are only considered if they geometrically could have the full number of hits, taking the limited coverage of the endplate into account. No further fiducial cuts were applied. After these cuts, the efficiency to reconstruct a hit is close to \SI{100}{\percent} for nearly all rows. The drop in efficiency to about \SI{96.6}{\percent} at row~27 is at the transition from readout module 0 to module 3, as labelled in figure~\ref{sfig:endplateNumbered}. Here the distance between both modules is about \SI{4}{\mm}, which causes distortions in the electric field. The distortions lead to a loss of charge and to a smaller hit-finding efficiency. Between module 3 and 5 (see figure \ref{sfig:endplateNumbered} for the numbering scheme), the distance is only about \SI{3}{\mm}. This leads to smaller field distortions and a much reduced loss in hit finding efficiency.

\subsection{Drift Velocity}
\begin{figure}[tb]
\begin{subfigure}[b]{0.48\textwidth}
\iftoggle{blackandwhite}{\includegraphics[width=\textwidth]{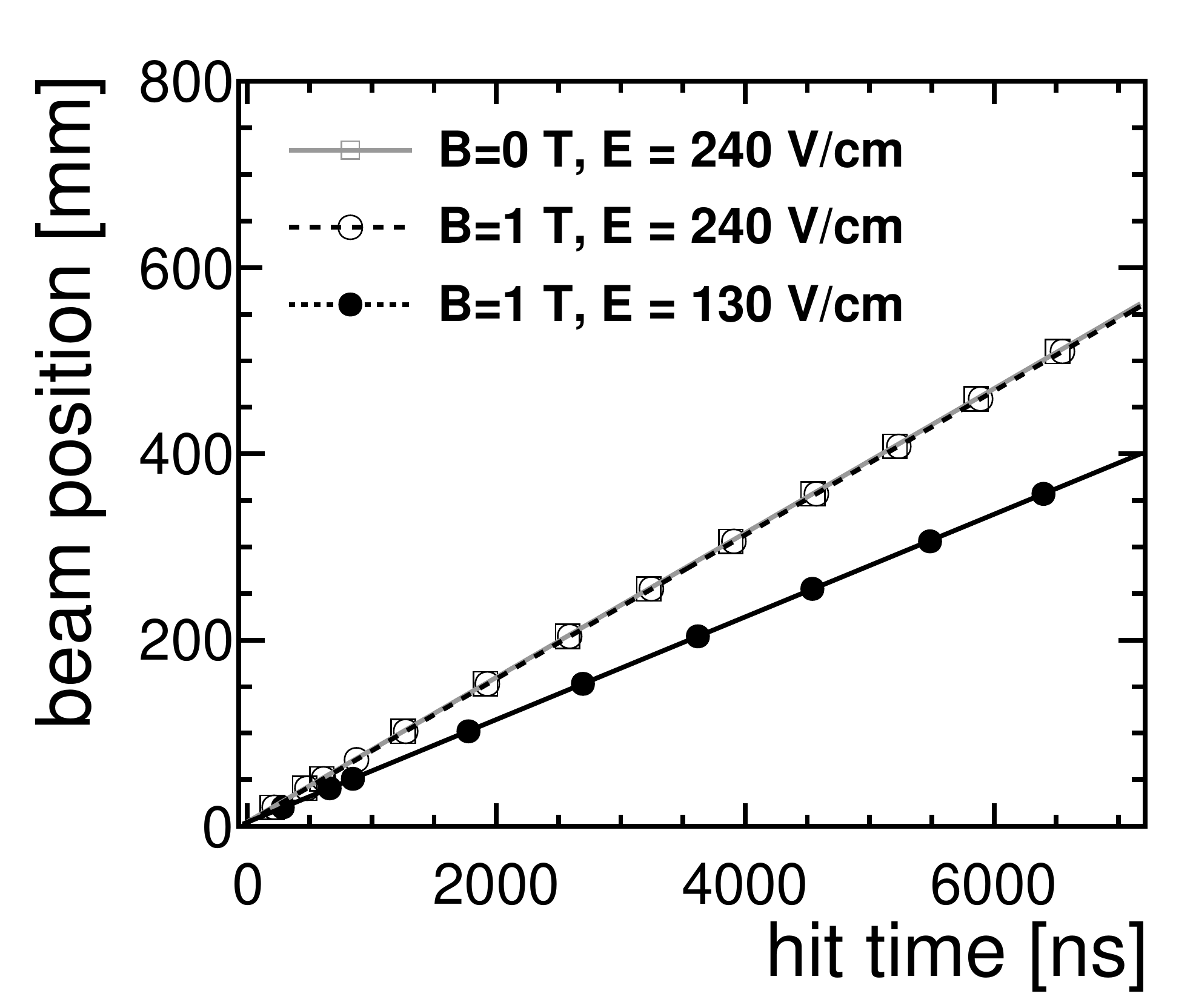}}{\includegraphics[width=\textwidth]{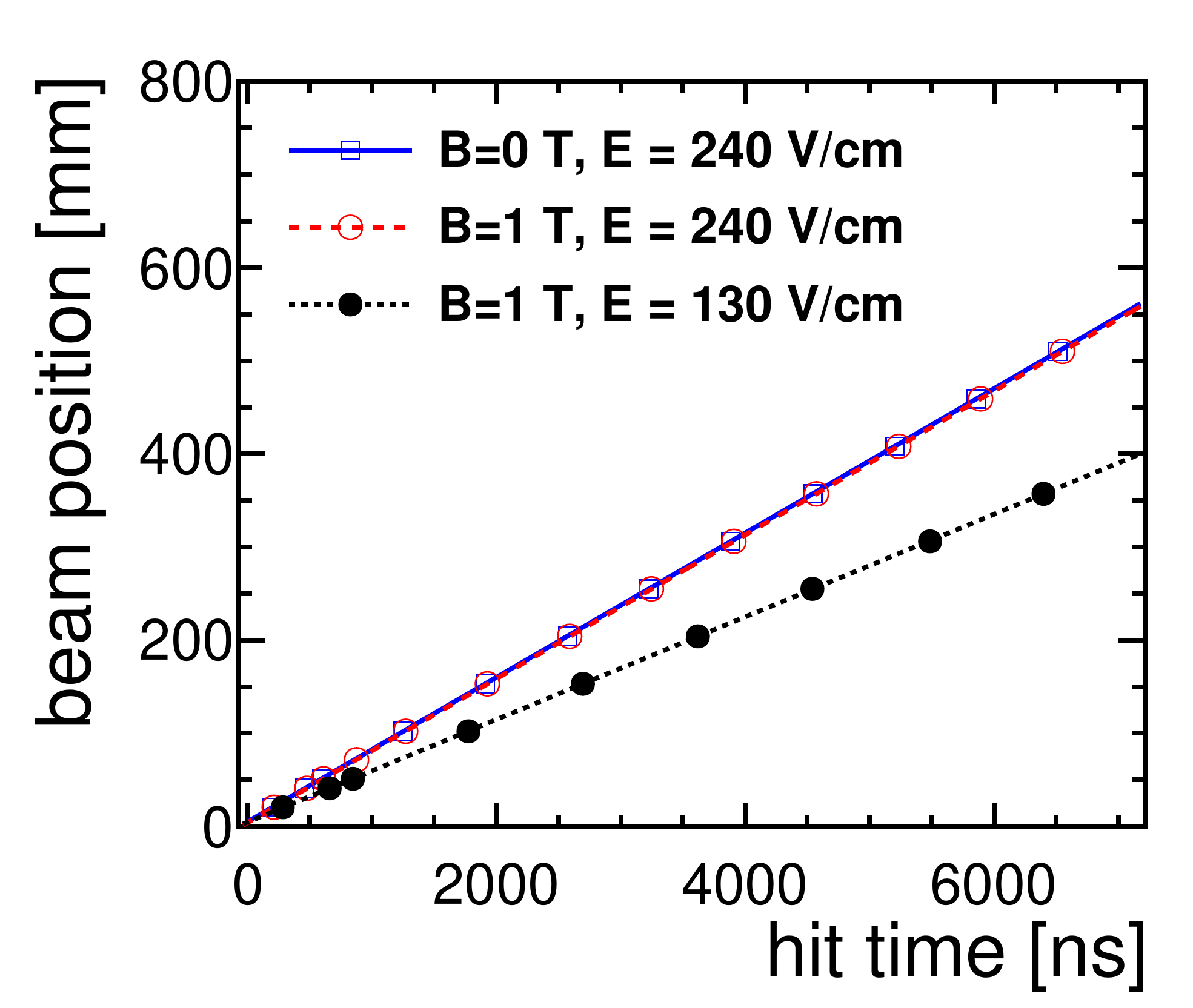}}
\caption{}
\label{sfig:vdrift}
\end{subfigure}
\hfill
\begin{subfigure}[b]{0.48\textwidth}
\iftoggle{blackandwhite}{\includegraphics[width=\textwidth]{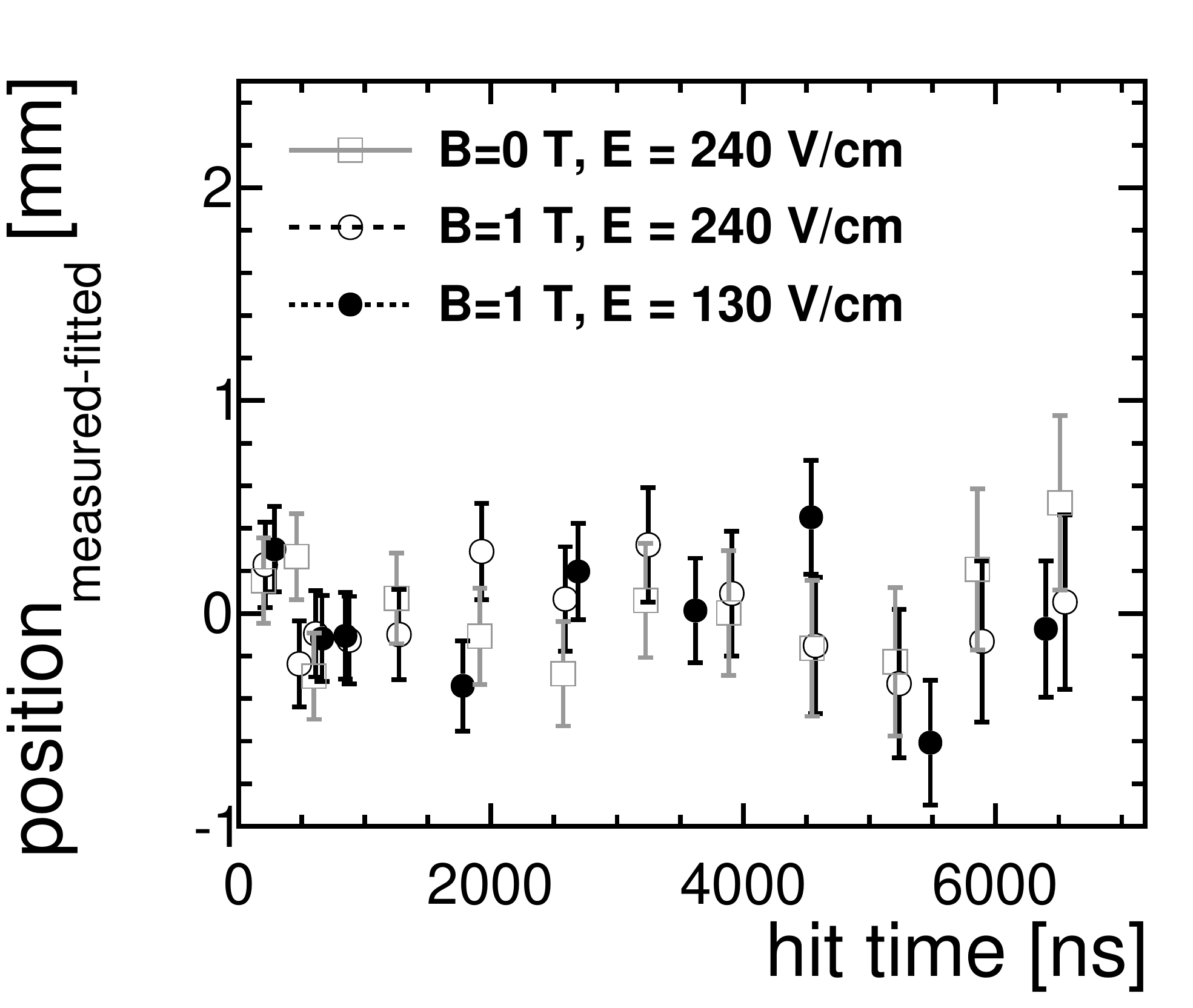}}{\includegraphics[width=\textwidth]{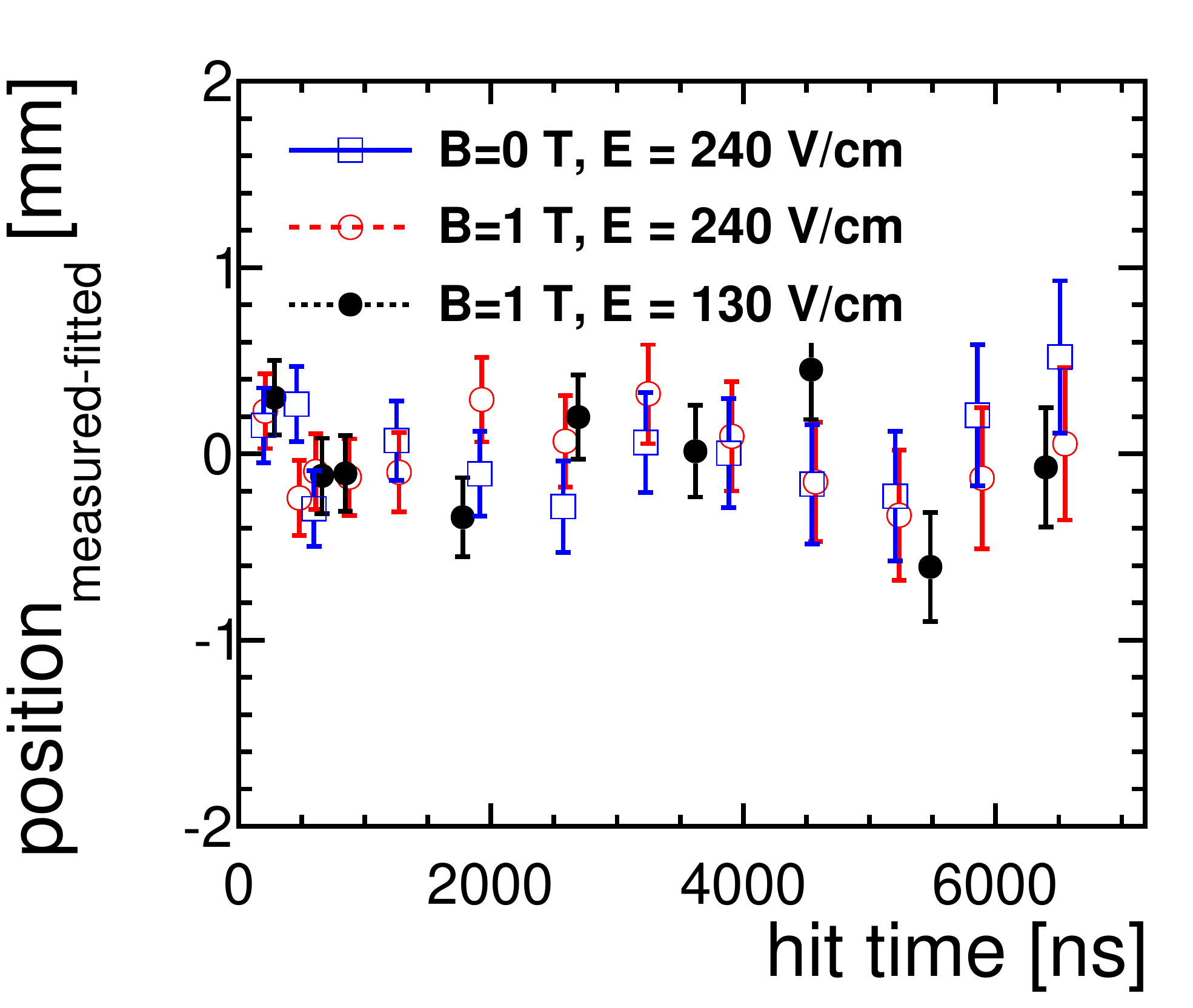}}
\caption{}
\label{sfig:vdriftDiff}
\end{subfigure}
\caption{\small \protect\subref{sfig:vdrift})~Beam position versus reconstructed time of the hit signal for \SI{0}{\tesla} and \SI{1}{\tesla} for the default drift field of \SI{240}{V/cm} and for a lower drift field of \SI{130}{V/cm} at \SI{1}{\tesla}. \protect\subref{sfig:vdriftDiff})~Deviation of each measured beam position from the fitted line through all measured beam positions for the three measurement runs.}
\label{fig:vdrift}
\end{figure}

The drift velocity was determined by moving the stage with the setup in several well-defined steps along the drift direction and reconstructing the position of the beam for each step. Data was taken for two different drift fields, as well as with and without magnetic field. The results are shown in figure~\ref{sfig:vdrift}. A straight line was fit to the data points of each measurement to determine the drift velocity from its slope. The start of the measurement volume corresponds to the intersection point of the lines. Figure \ref{sfig:vdriftDiff} shows the deviation of the measured beam positions from the line resulting from the fit. The errors of the measurements come mainly from the uncertainty in the position of the stage and to a small part from the intrinsic accuracy of the reconstruction. 
The resulting drift velocities are listed in table~\ref{tab:drift}, where they are also compared to the expectations from Magboltz~\cite{Magboltz} simulation and show a good agreement.

\begin{table}[hbt]
\centering
\begin{tabular}{ll|l|l}
\hline
E-field  [\si[per-mode=symbol]{\V \per \cm}] & B-Field [\si{\tesla}]& $v_{\rm drift} [\si[per-mode=symbol]{\um \per \ns}]$ & $v_{\rm drift}^{\rm sim} [\si[per-mode=symbol]{\um \per \ns}]$\\
\hline
240 & 0 & $77.52 \pm 0.06$ & $77.02$ \\
240 & 1 & $77.26 \pm 0.04$ & $76.95$ \\
130 & 1 & $55.12 \pm 0.03$ & $55.63$ \\
\hline
\end{tabular}
\caption{\small\label{tab:drift} Drift velocities for different electric and magnetic fields. The two drift fields chosen correspond to the maximum of the drift velocity versus drift field relation (\SI[per-mode=symbol]{240}{\V \per \cm}) and to the point of minimal transverse diffusion (\SI[per-mode=symbol]{130}{\V \per \cm}) for the gas used. The simulation has been done using Magboltz~\cite{Magboltz}.}
\end{table}

\subsection{Point Resolution}
Once the drift velocity is known, the hit time can be converted into a spatial coordinate, to complete the three-dimensional hit coordinate. The point resolution is calculated in the GEM plane along the pad rows ($r\varphi$), and perpendicular to the GEM plane along the drift direction ($z$). The resolution is determined from the width of the residual distribution. The residuals are defined in the $r\varphi$ plane as the distance between the hit and the reconstructed track along the pad row. In the $z$ direction, the residuals are defined as the equivalent distance along the drift direction, perpendicular to the readout plate. To get an unbiased estimate of the resolution, the residuals are calculated both from a track fit including the hit under study, as well as a track fit excluding this hit. The best estimate of the resolution is then the geometric mean of the widths of these two distributions~\cite{resolution}.

\begin{figure}[tb]
\begin{subfigure}[b]{0.48\textwidth}
\iftoggle{blackandwhite}{\includegraphics[width=\textwidth]{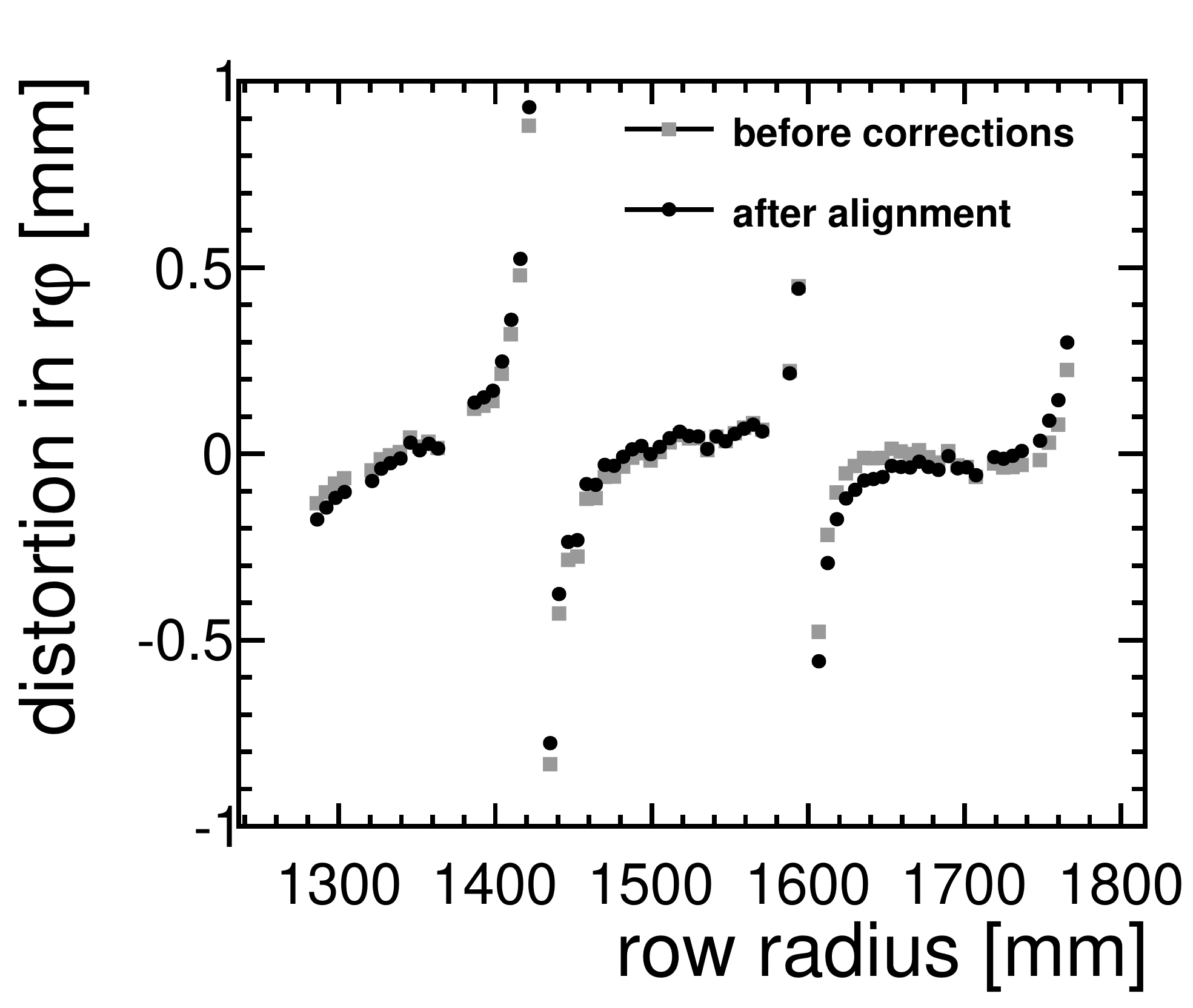}}{\includegraphics[width=\textwidth]{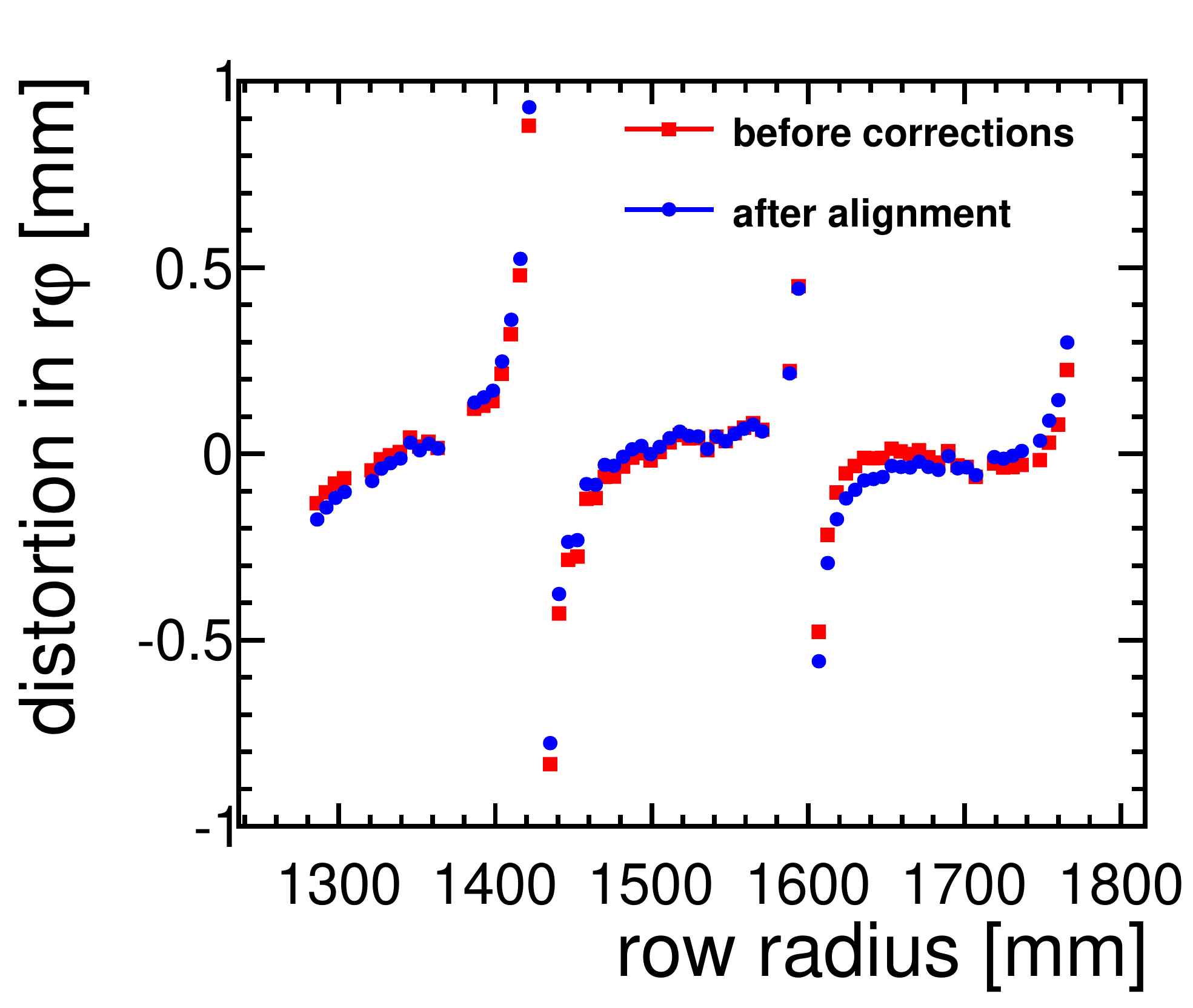}}
\caption{}
\label{sfig:1Talign}
\end{subfigure}
\hfill
\begin{subfigure}[b]{0.48\textwidth}
\iftoggle{blackandwhite}{\includegraphics[width=\textwidth]{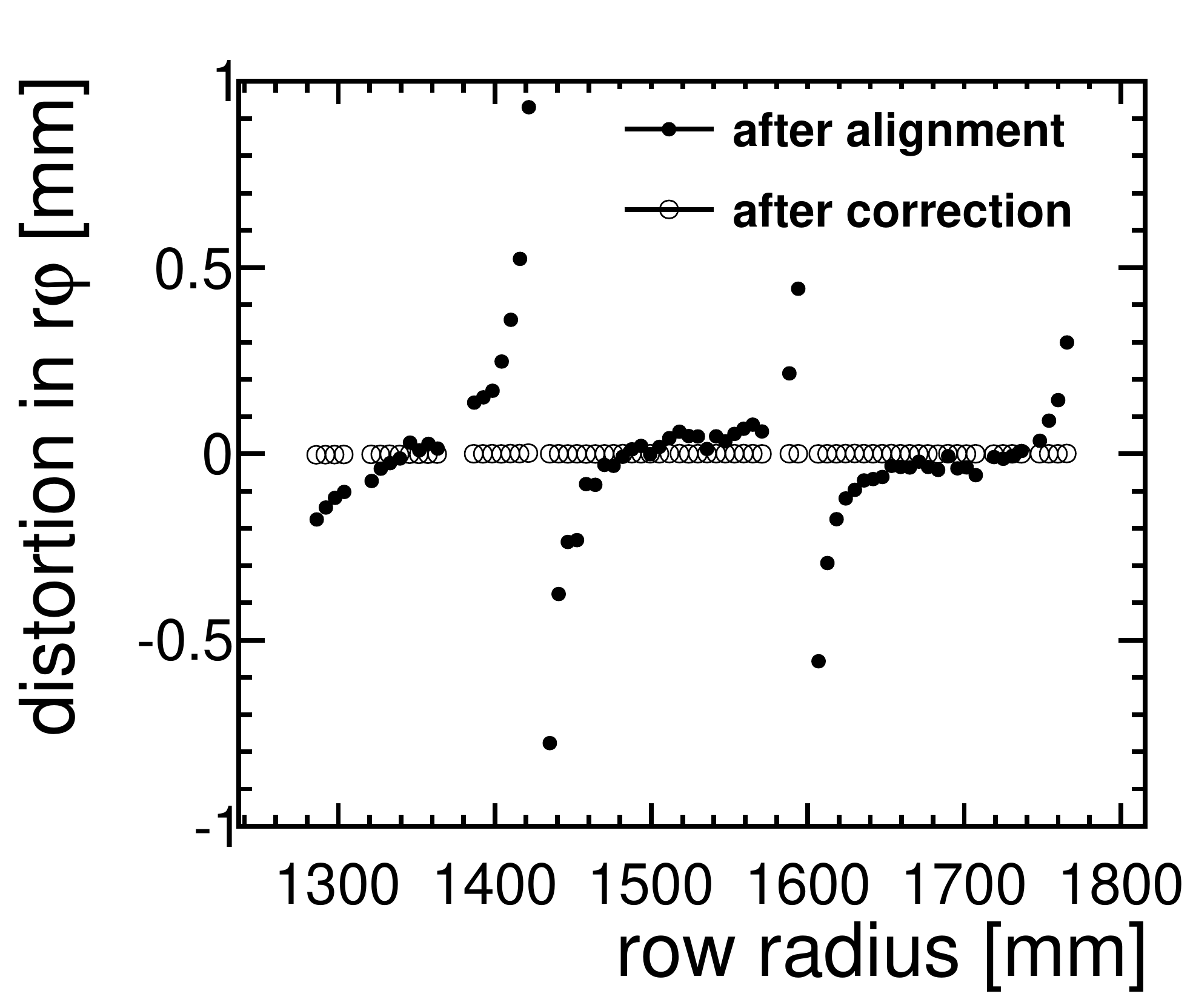}}{\includegraphics[width=\textwidth]{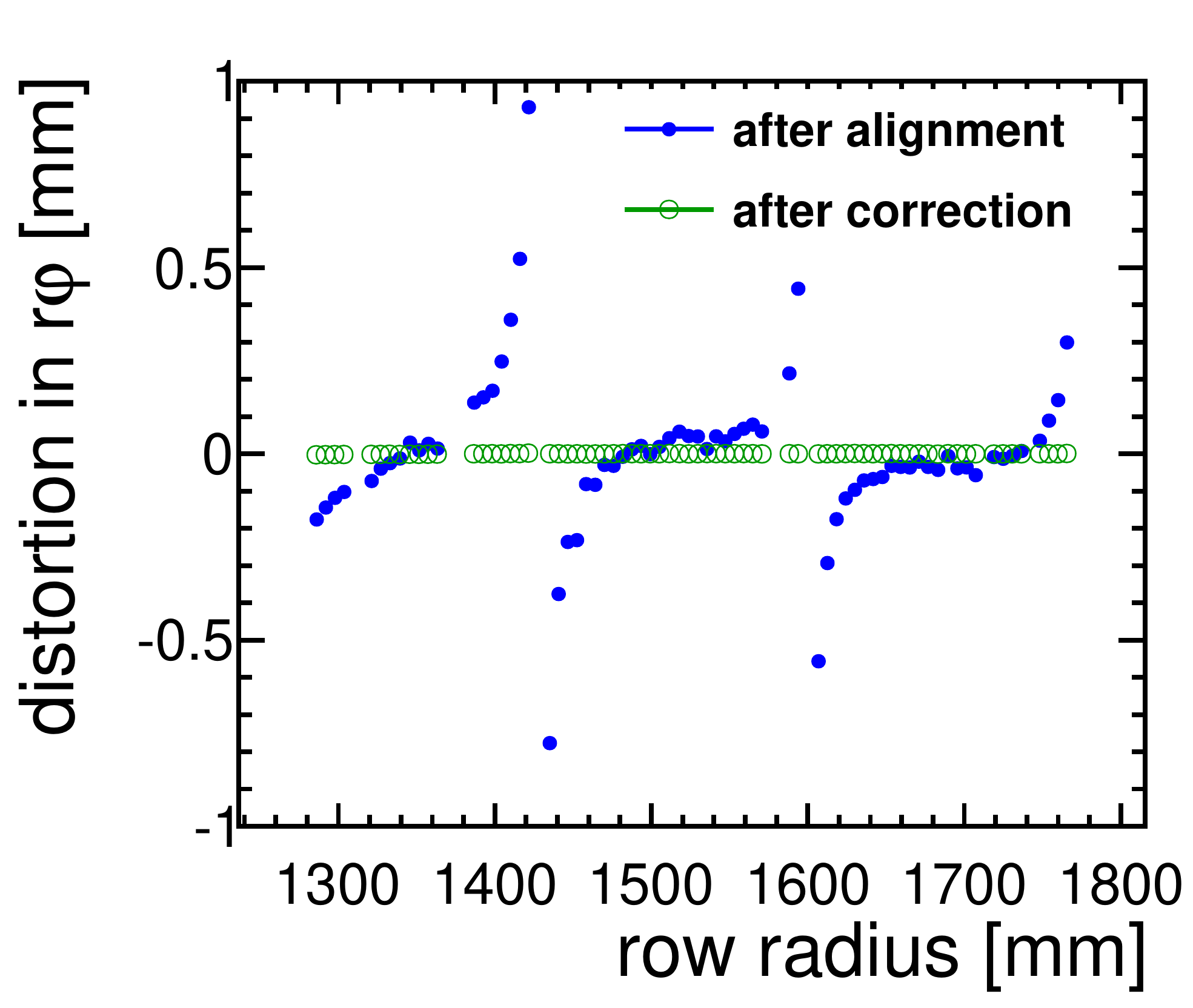}}
\caption{}
\label{sfig:1Tdistort}
\end{subfigure}
\caption{\small Mean hit position in $r\varphi$ with respect to the track position at \SI{1}{\tesla} versus pad row radius. \protect\subref{sfig:1Talign})~Alignment correction. \protect\subref{sfig:1Tdistort})~Distortion correction.}
\label{fig:align1Tdistort}
\end{figure}

The mean of the distribution of the residuals clearly shows strong systematic effects, (see figure \ref{fig:align1Tdistort}), 
which are due to distortions and misalignment. Distortions are caused by non-perfect electric fields in particular close to boundaries between modules. They result in large biases of hits close to the boundaries. In addition, modules can be shifted and rotated relative to their nominal position. This misalignment results in linear displacements of the hits within one module.

To obtain the ultimate performance of the system, these effects need to be corrected for. 
The corrections are obtained in an iterative process. Alignment is accounted for by overall rotations and shifts which are determined on a module-by-module basis. Distortions are accounted for by systematic offsets, which are determined for each row individually. They are most pronounced at the edges of the module.

\begin{figure}[tb]
\centering
\iftoggle{blackandwhite}{\includegraphics[width=0.85\textwidth]{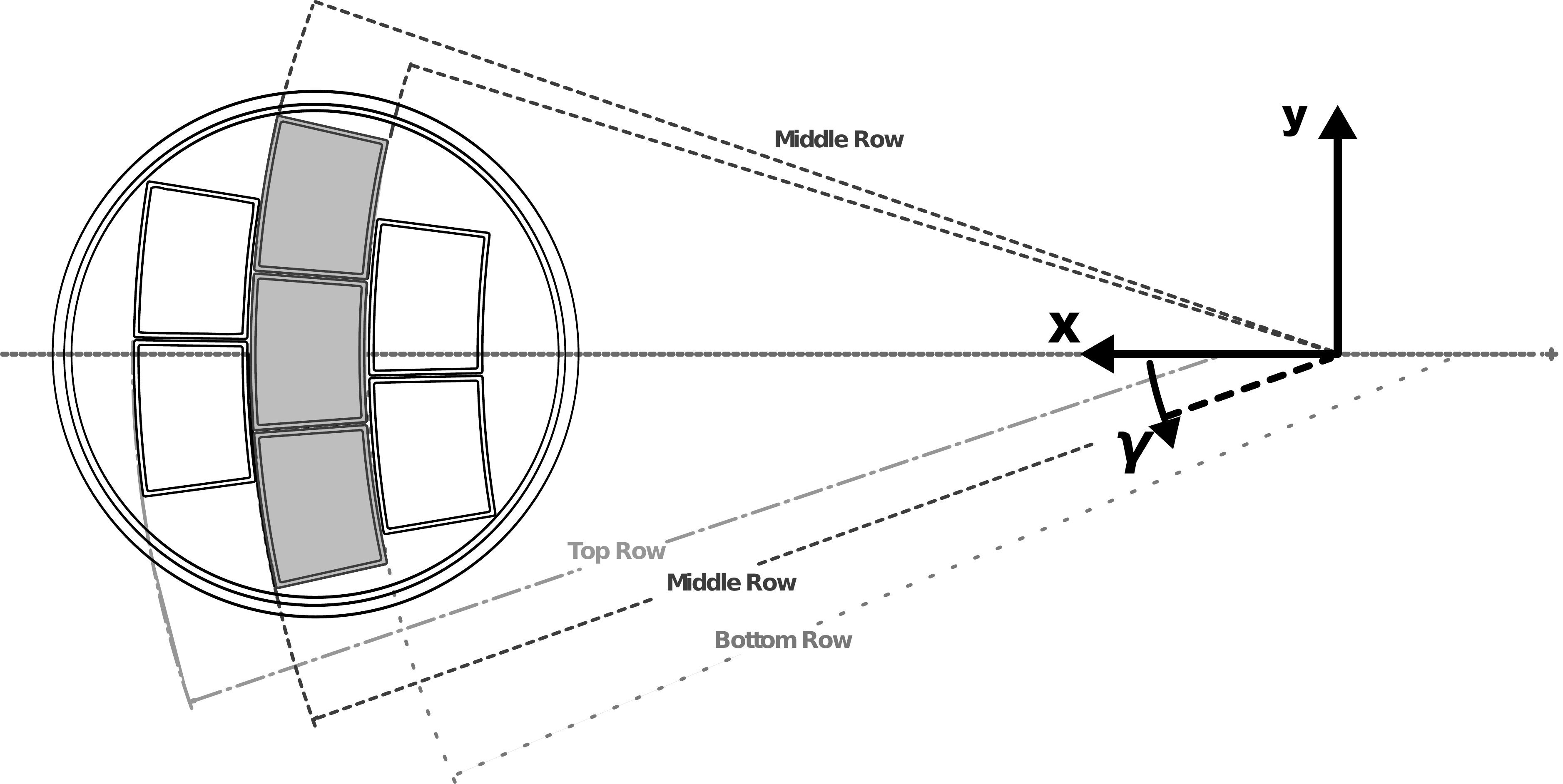}}{\includegraphics[width=0.85\textwidth]{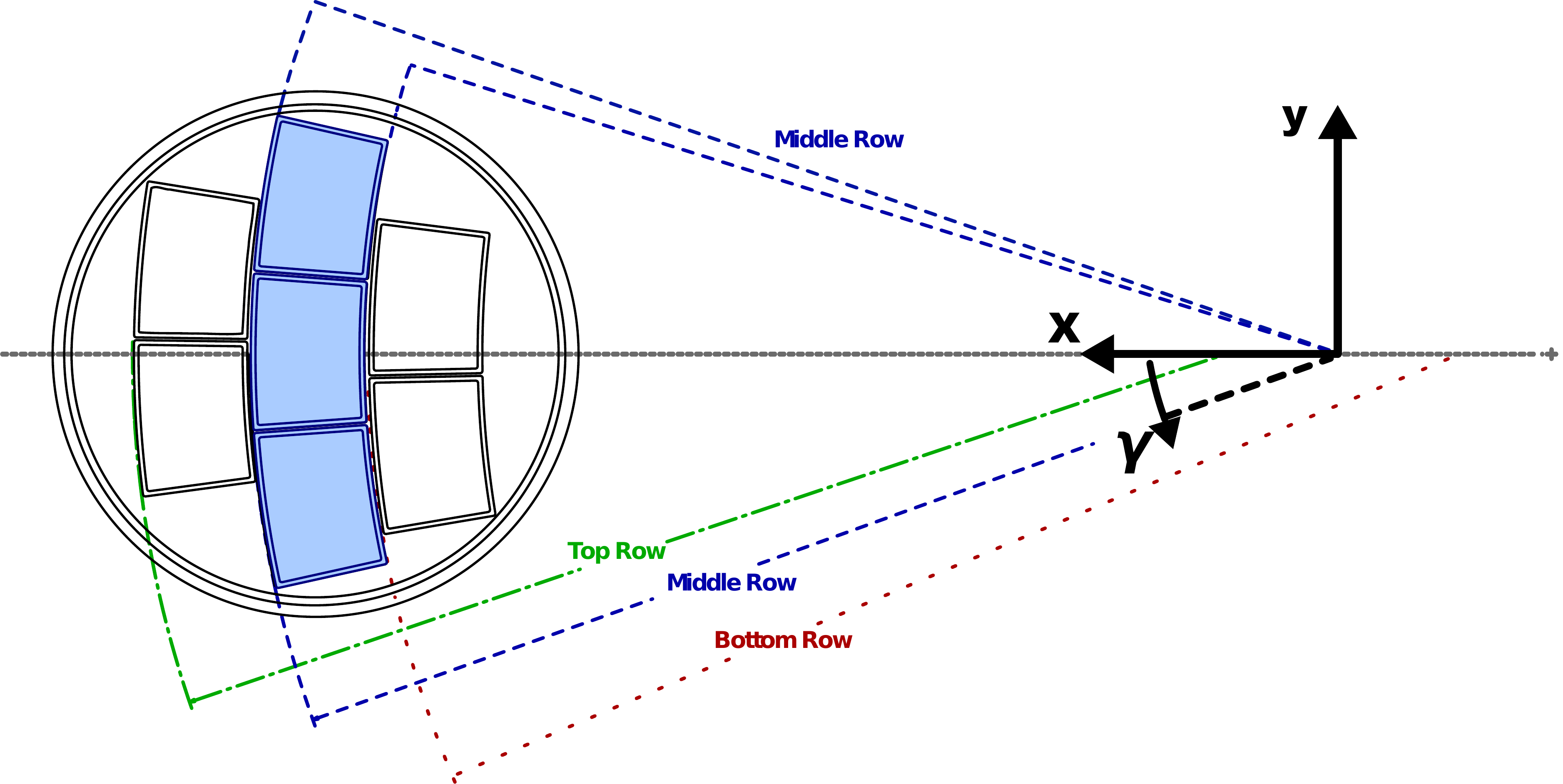}}
\caption{\small Coordinate systems used to determine the alignment parameters for the modules as explained in the text. The coordinate axis x, y and the rotation $\gamma$ are drawn for the central module row (modules highlighted). The dotted horizontal line marks the radial line that crosses the centre of the endplate and defines the x-axis. The crossings of the different dashed lines define the centres of the coordinate systems for each module row.}
\label{fig:coordinate}
\end{figure}

The alignment parameters are determined for each module in a right-handed Cartesian coordinate system, which is the same as is implemented in the geometry description toolkit Gear~\cite{GearILCSoft,EudetGearTPC} of the reconstruction and analysis software. It has its origin in the centre of the circle described by the rows of the readout module, see figure~\ref{fig:coordinate}. The x-axis is defined as the radial line that crosses the centre of the endplate. The y-axis is perpendicular to it, parallel to the endplate. The z-axis is perpendicular to the endplate. As described in section \ref{sec:fieldcage}, the modules are arranged on an arc of the same radius in each row. Therefore, the centre of the alignment coordinate system is displaced for each module row along the x-axis by the pitch of the module row, so that the rotational displacement $\gamma$ describes a displacement along the arc of that module row. In the alignment, the choice of this coordinate system is mathematically equivalent to systems that 
have their origin in the centre of the respective module and describe rotations around this centre and translations along the x and y-axes. 

The parameters are determined in a two-step procedure. In a first step, a global fit to the hits is performed, with the alignment constants as free parameters. For each module, an offset in x and y as well as a rotation around the centre of its coordinate system, is allowed. The z coordinate is fixed since the measurement is not very sensitive in this direction. A multi-dimensional $\chi^2$-minimisation is performed, using the Millepede II toolkit to find the best set of parameters. To exclude influences from $\vec{E} \times \vec{B}$ effects, the minimisation is done simultaneously for all measurement runs taken at \SI{0}{\tesla}. 
The obtained alignment is applied to all measurement runs, with and without magnetic field. The alignment leads to corrections of the order of \SI{0.1}{\mm} along the x-axis and the y-axis and a few milliradians along the module row arc. The mean of the residuals before and after the alignment procedure are shown per pad row in figure~\ref{sfig:1Talign} for data taken in a $\SI{1}{\tesla}$ magnetic field.

In a second step, the distortions, i.e.\ the part of the systematic shifts in the residuals which cannot be explained by overall alignment constants, are determined. They are derived per row from the mean shift of the residuals, after alignment, and applied as corrections to the data. To ensure statistical independence, the residuals are determined on a sub-sample of the available data, the rest of the data is used to measure the effect of the correction. The effect of the distortion correction is shown in figure~\ref{sfig:1Tdistort}. It is visible that the systematic shifts of the residuals are close to zero after this step.

\begin{figure}[b!]
\begin{subfigure}[b]{0.48\textwidth}
\iftoggle{blackandwhite}{\includegraphics[width=\textwidth]{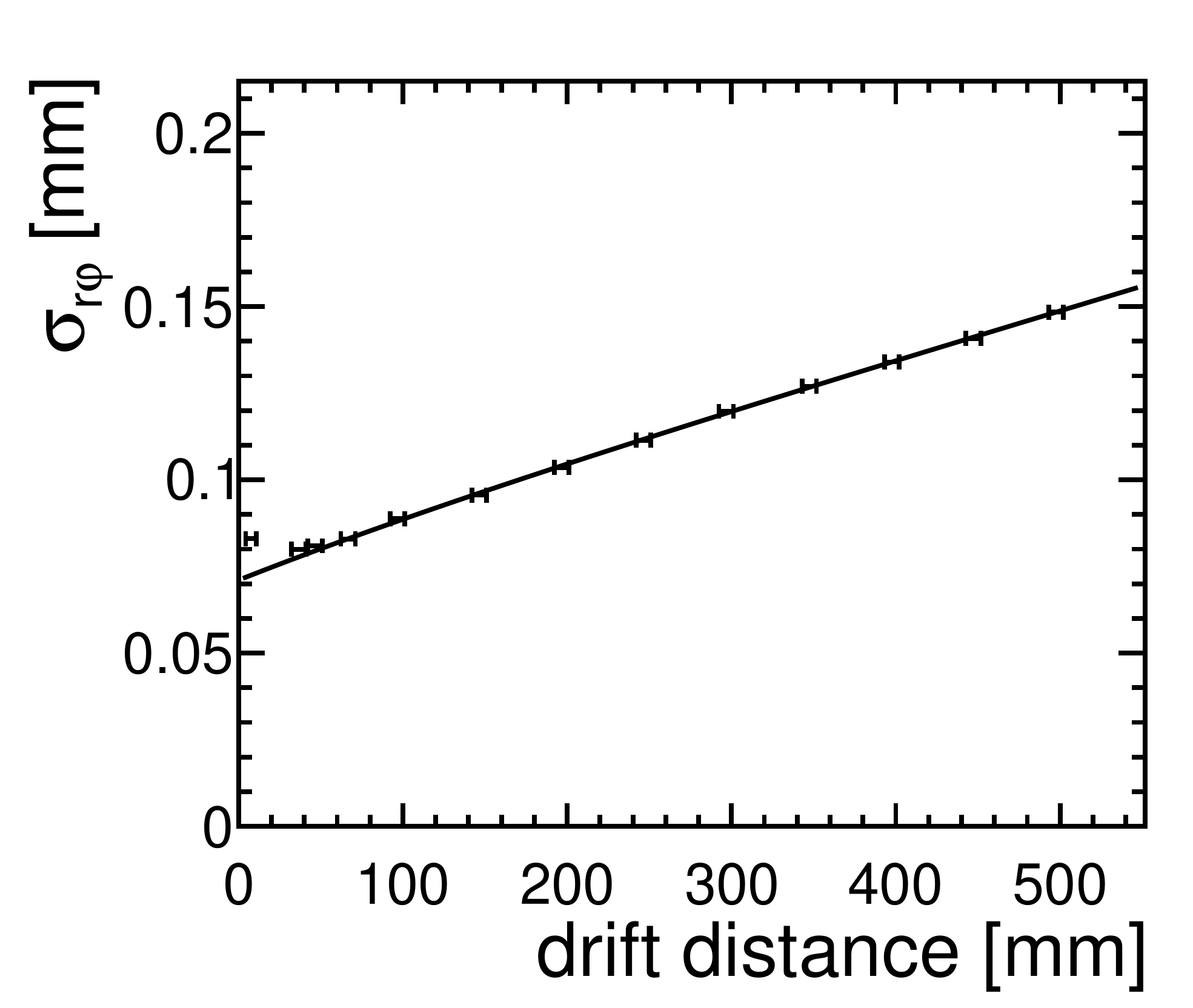}}{\includegraphics[width=\textwidth]{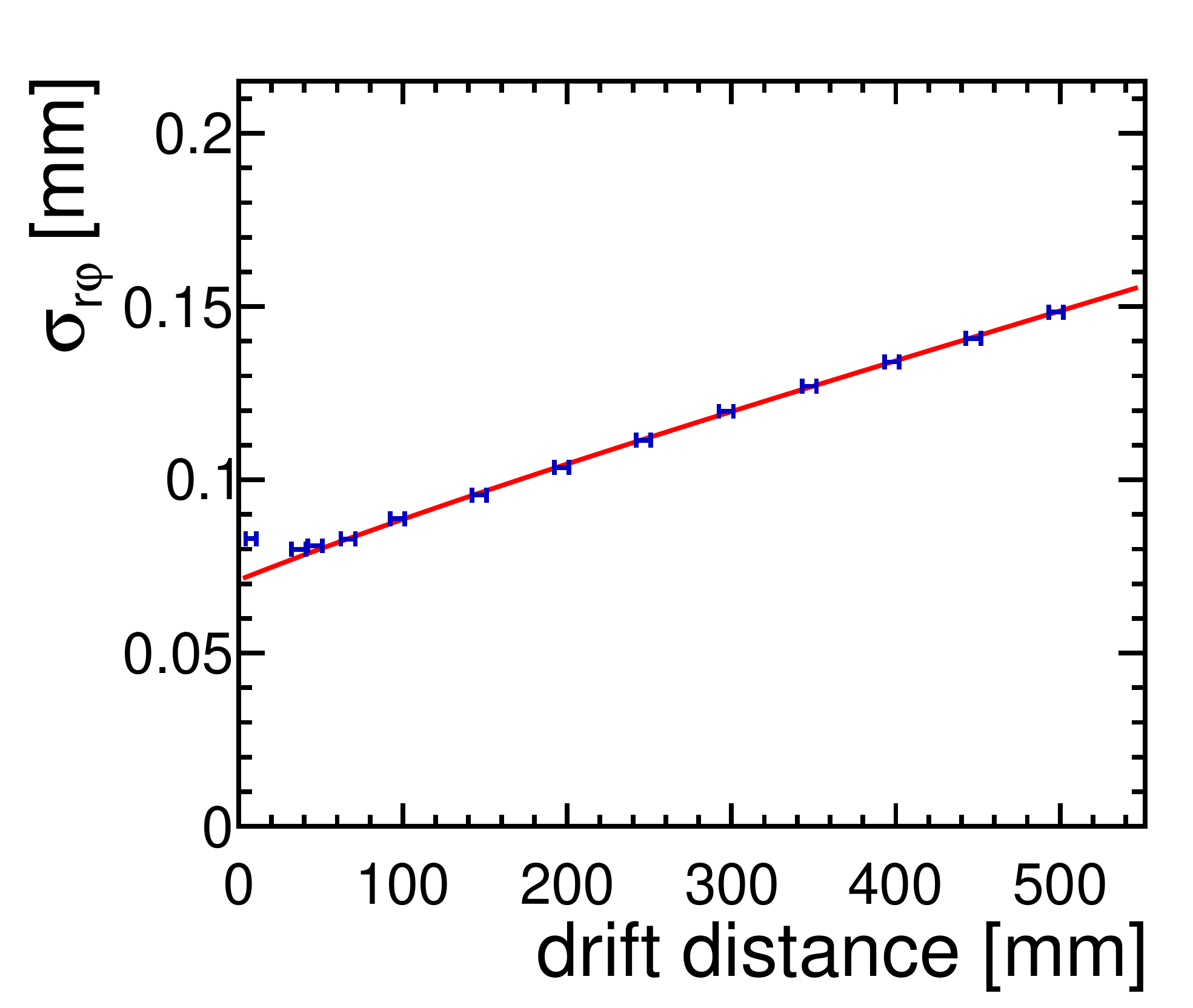}}
\caption{}
\label{sfig:resrphi}
\end{subfigure}
\hfill
\begin{subfigure}[b]{0.48\textwidth}
\iftoggle{blackandwhite}{\includegraphics[width=\textwidth]{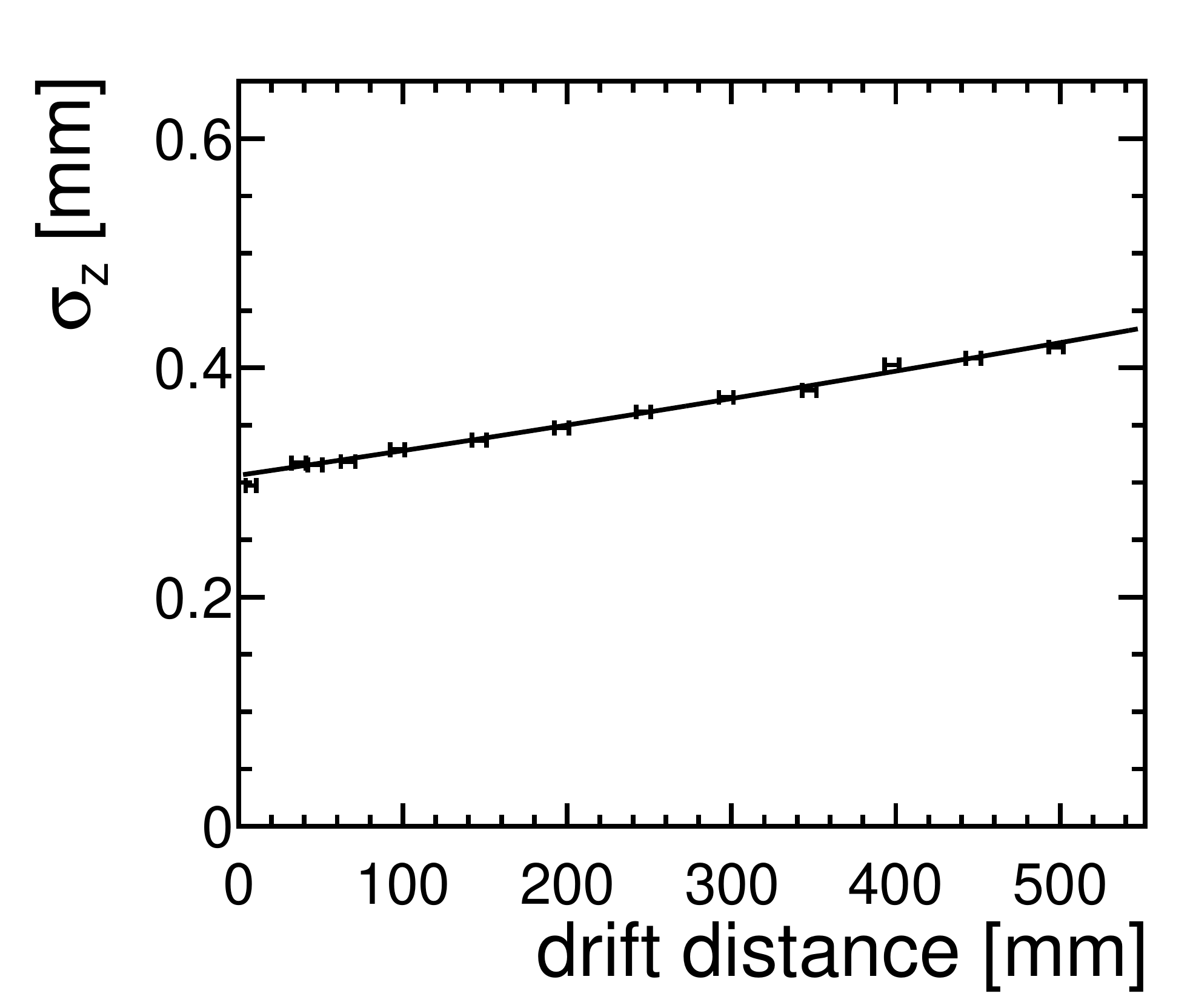}}{\includegraphics[width=\textwidth]{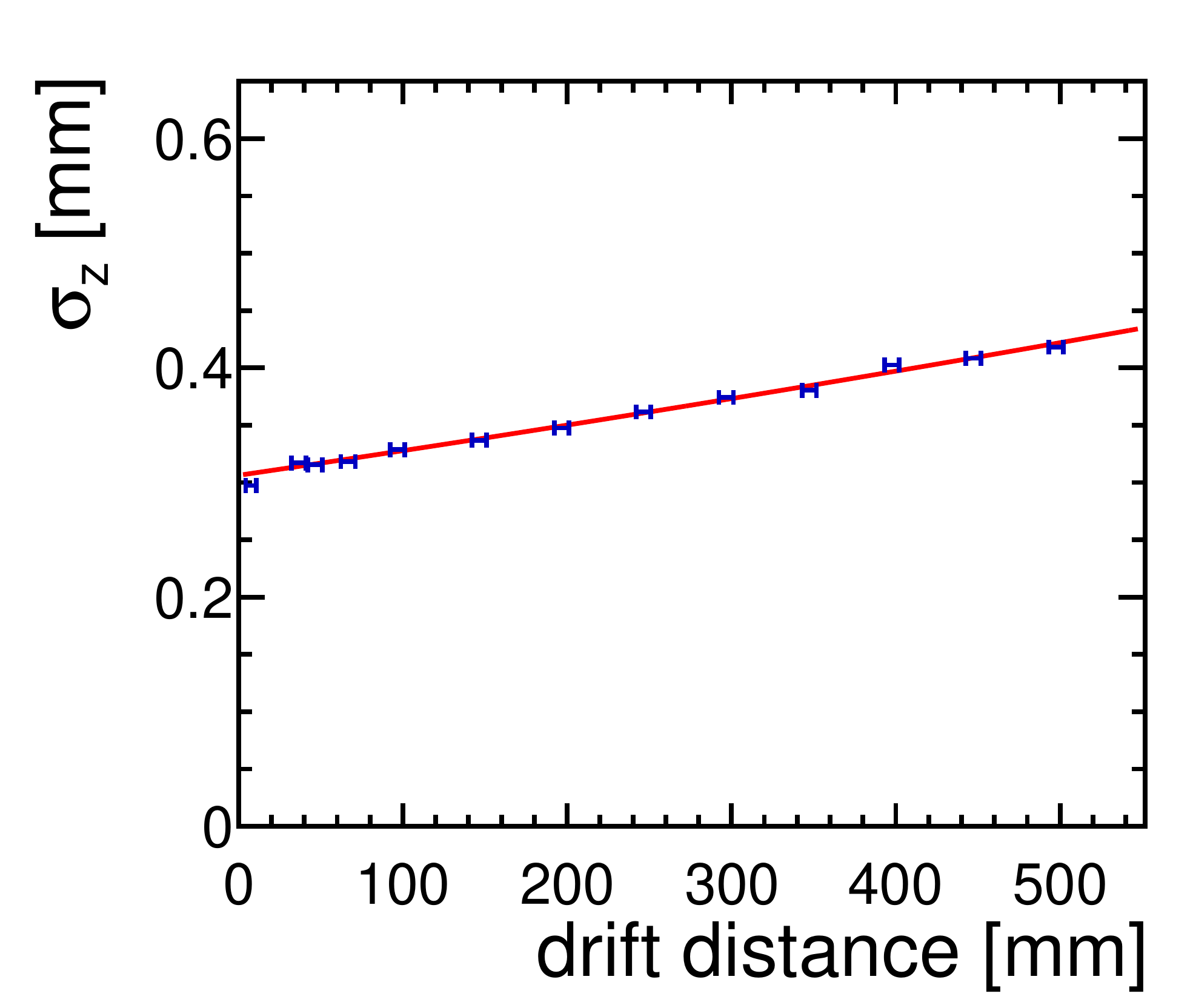}}
\caption{}
\label{sfig:resz}
\end{subfigure}
\caption{\small Point resolution depending on drift length for measurements at \SI{1}{\tesla} magnetic field \protect\subref{sfig:resrphi})~in $r\varphi$, \protect\subref{sfig:resz})~in $z$. The fit results are listed in table \ref{tab:res}.}
\label{fig:resolution}
\end{figure}

After all corrections have been applied, the widths of the distributions of the residuals are used to calculate the point resolution as described above, on a row-by-row basis. The mean of all rows as a function of the drift distance is shown in figure~\ref{sfig:resrphi} for the $r\varphi$ direction and in figure~\ref{sfig:resz} for the $z$ direction. 

In both plots, the following function, describing the point resolution $\sigma_{r\varphi/z}(z)$ in $r\varphi$ and $z$, respectively, as a function of the drift distance for tracks that run perpendicular to the pad rows~\cite{Blum:2008}, has been fitted to the data points:
\begin{equation}
\sigma_{r\varphi/z} (z) =\sqrt{\sigma_{0,r\varphi / z}^2+\frac{D_{t/l}^2}{N_{\mathrm{eff}}\cdot e^{-Az}}z} \quad . 
 \label{math:zResolutionElectronLoss}
\end{equation}
Here, $\sigma_{0,r\varphi / z}$ describes the intrinsic resolution in $r\varphi$ and $z$, respectively, of the readout at zero drift distance. For B=\SI{1}{\tesla} and E=\SI[per-mode=symbol]{240}{\V\per\cm}, the longitudinal diffusion $D_{l}=\SI[per-mode=symbol]{0.226}{\mm \per \sqrt{\cm}}$ was derived from a Magboltz simulation. The transverse diffusion $D_{t}$ was determined from data from the measured width of the pad response function (PRF)~\cite{Blum:2008}. The PRF describes the average signal shape measured along the pads in a row. Its width depends, among other parameters, on the diffusion of the charge cloud. A fit of the function $w(z) = \sqrt{w(0) + D_t^2 z}$ to the measured PRF widths $w$ at different drift lengths $z$ results in a value of $\SI[per-mode=symbol]{0.1032}{\mm \per \sqrt{\cm}}$ with a statistical error of $\SI[per-mode=symbol]{0.0004}{\mm \per \sqrt{\cm}}$ for $D_{t}$. For 
the fit of function \eqref{math:zResolutionElectronLoss}, 
the central value of $D_t$ is used.

$N_{\mathrm{eff}}$ describes the effective number of signal electrons contributing to the measurement~\cite{YonamineJinst2014}. The term $e^{-Az}$ describes the loss of signal electrons during the drift due to attachment to gas molecules, primarily oxygen impurities, with the attachment factor $A$ being a free fit parameter. 

The fit is performed with $\sigma_{0}$, $N_{\mathrm{eff}}$ and $A$ as free parameters.  The results are listed in table~\ref{tab:res}. For very short drift distances, the charge cloud size becomes similar to the pad pitch, resulting in a deterioration of the transverse resolution. Therefore, only measurements with a drift distance larger than \SI{70}{\mm} are included in the fit. 
Within errors, the results of $N_{\mathrm{eff}}$ and the attachment rate $A$ are equal for the fits of the longitudinal and transverse resolution.
The values of 39.8 and 39.5 for $N_{\mathrm{eff}}$ are similar to the result of a Heed~\cite{Heed} 
simulation. The results for the intrinsic resolution $\sigma_0$ and attachment rate $A$ are comparable to the results from measurements in 2012 described in~\cite{Kato:2014uha} with a different GEM-based module.

\begin{table}[bt]
\centering
\begin{tabular}{l|l|l|l|l}
\hline
$\sigma$ & $\sigma_{0,r\varphi / z} [\si{\um}] $ & $N_{\mathrm{eff}}$ & $A [\si{m^{-1}}]$ & $D_{t/l} [\si[per-mode=symbol]{\mm \per \sqrt{\cm}}]$ (fixed) \\
\hline
$r\varphi$ &  71.0 $\pm$ 1.2 & 39.8 $\pm$ 2.0 & 0.495 $\pm$ 0.097 & 0.103 \\
$z$     & 306.3 $\pm$ 0.8 & 39.5 $\pm$ 1.6 & 0.529 $\pm$ 0.084 & 0.226\\
\hline
\end{tabular}
\caption{\small \label{tab:res} Table of results for the fit of equation \eqref{math:zResolutionElectronLoss} to the measured point resolution in $r\varphi$ and $z$ shown in figure~\ref{fig:resolution}. $D_t$ and $D_l$ are given for a drift field of \SI[per-mode=symbol]{240}{\V\per\cm} and a magnetic field of \SI{1}{\tesla}.}
\end{table}

\begin{figure}[b!]
\centering
\iftoggle{blackandwhite}{\includegraphics[width=0.48\textwidth]{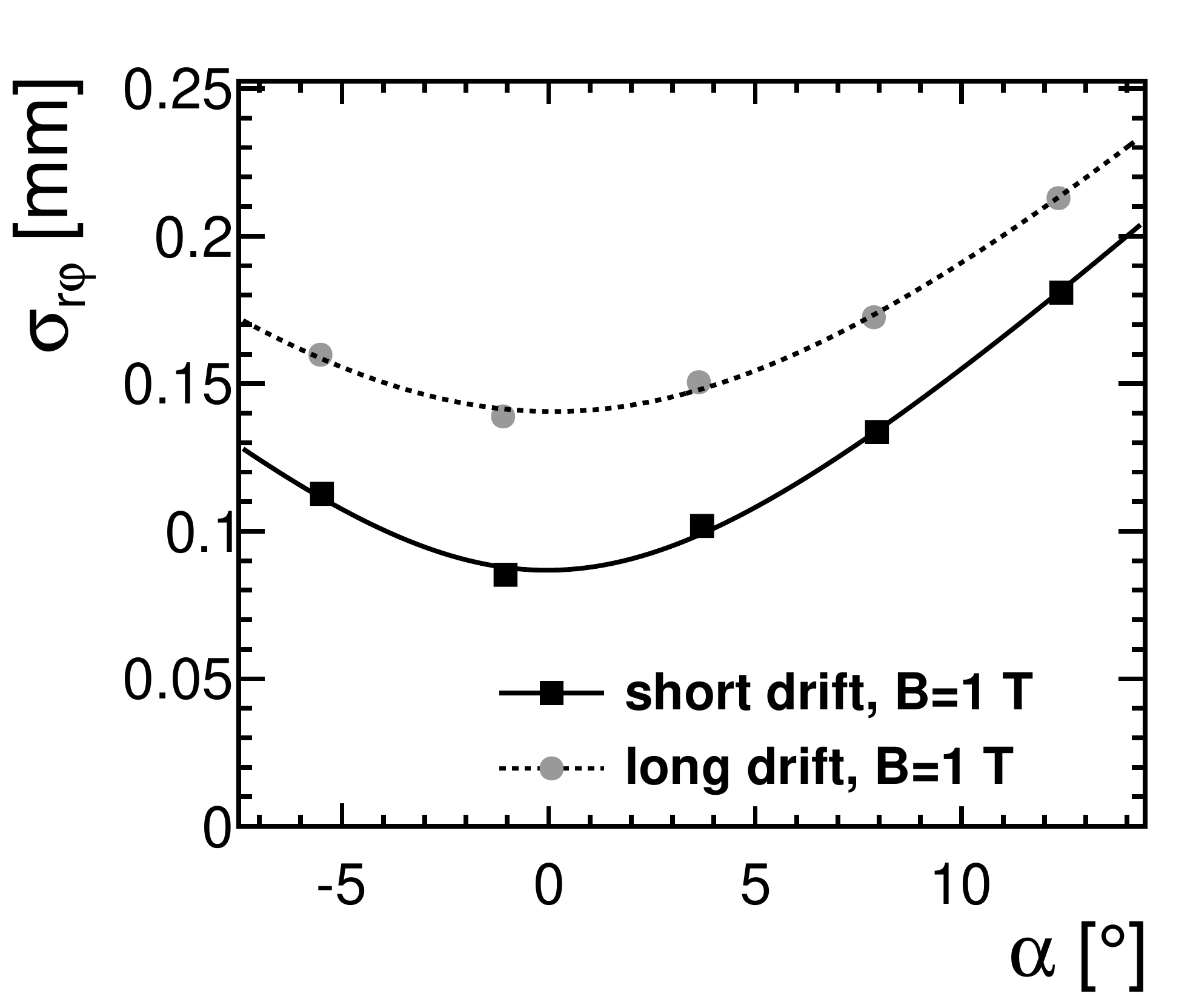}}{\includegraphics[width=0.48\textwidth]{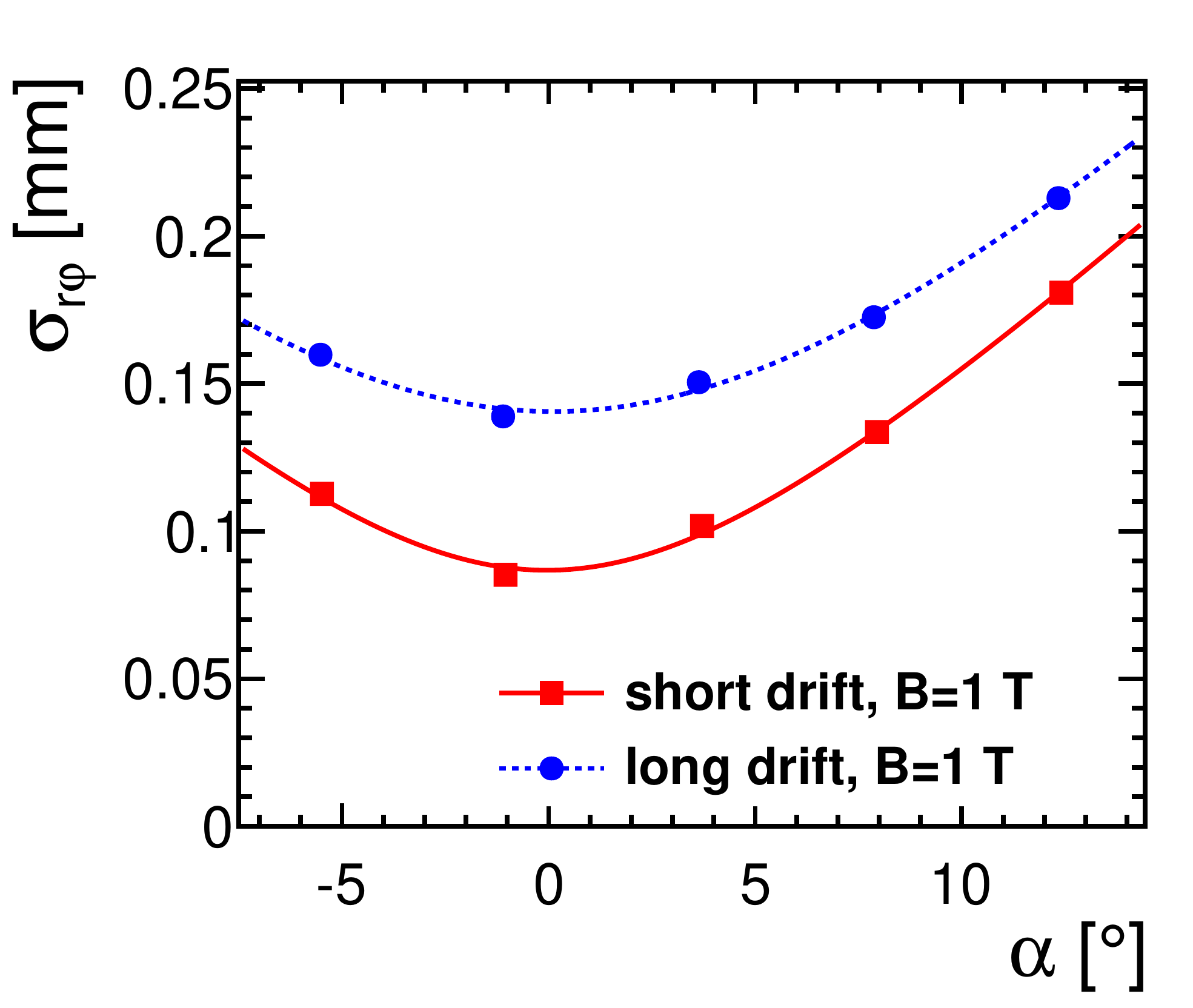}}
\caption{\small Point resolution depending on azimuthal track angle $\alpha$ with respect to the normal to the pad row, measured at \SI{1}{\tesla} magnetic field. Plotted for a drift distance of \SI{10}{\cm} (short drift) and \SI{40}{\cm} (long drift), including fits of equation \eqref{math:phiResolution}.}
\label{fig:restheta}
\end{figure}

Following \cite{Blum:2008,YonamineJinst2014}, the resolution formula for inclined tracks has to take into account the azimuthal angle $\alpha = {\varphi}_{pad} - {\phi}_{track}$ with respect to the normal to the pad row. In the limit of small angles $\alpha$, the resolution can be approximated as
\begin{equation}
  \sigma_{r\varphi}(z,\alpha) \approx \sqrt{\sigma_{r\varphi}^2(z)+\frac{L^2}{12\hat{N}_{\mathrm{eff}}}\tan^2\alpha} \quad .\\
 \label{math:phiResolution}
\end{equation}
Here, $\hat{N}_{\mathrm{eff}}$ is the effective number of clusters collected over the height $L$ of a pad row.
Figure~\ref{fig:restheta} shows the measured point resolution in $r\varphi$ as a function of the azimuthal angle $\alpha$. For this plot, tracks at a drift distance of \SI{10}{\cm} and \SI{40}{\cm} have been selected. The dependence on the azimuthal angle shows the behaviour with $\tan \lvert \alpha \rvert$ as expected from equation \eqref{math:phiResolution}.

\begin{figure}[tb]
\centering
\iftoggle{blackandwhite}{\includegraphics[width=0.48\textwidth]{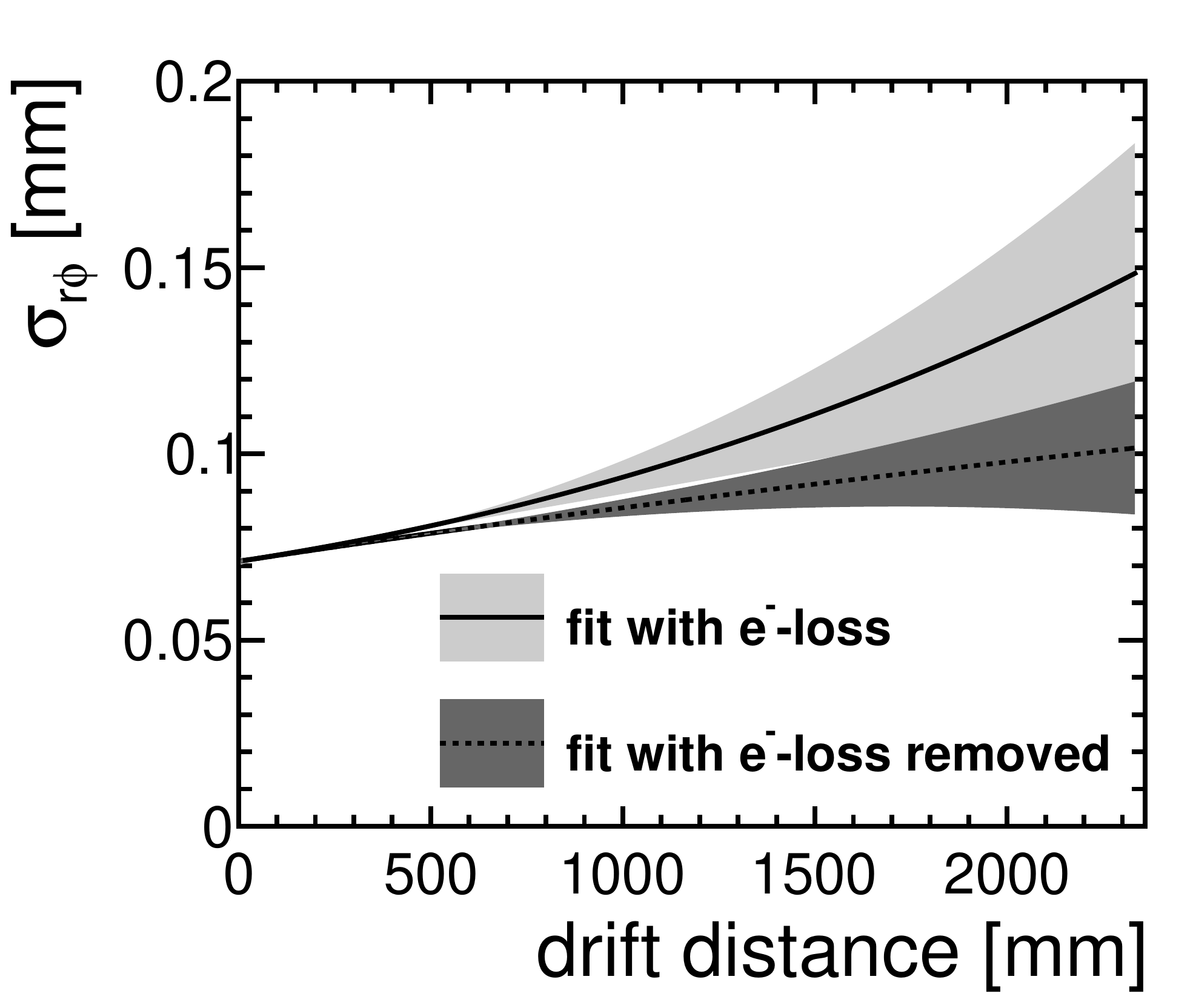}}{\includegraphics[width=0.48\textwidth]{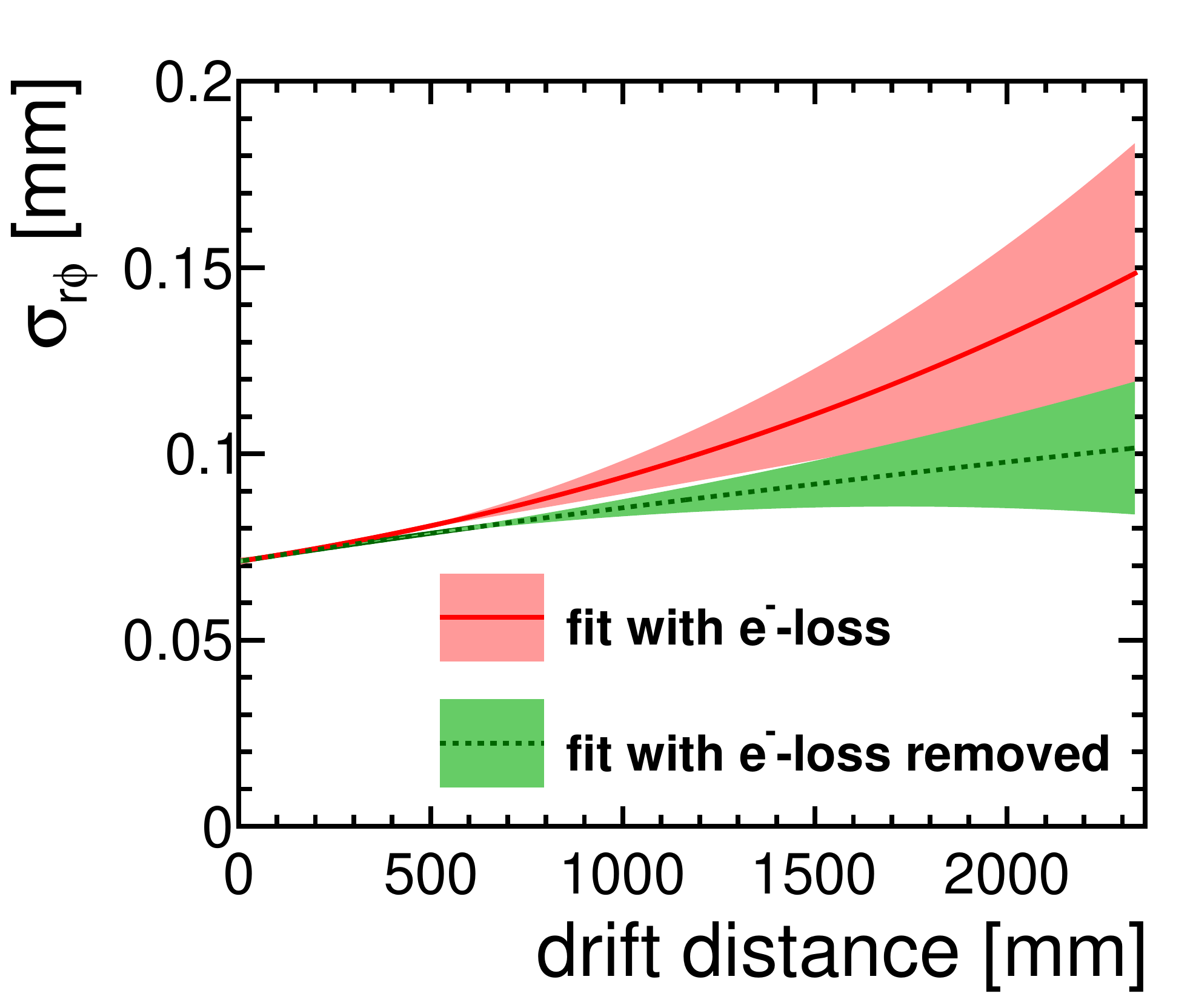}}
\caption{\small Extrapolation of the point resolution to a magnetic field of \SI{3.5}{\tesla} and plotted over the full ILD TPC drift length of \SI{2.35}{\m} including $1\sigma$ error bands. In the upper curve with the measured attachment rate, in the lower curve without any attachment.}
\label{fig:resextrapol}
\end{figure}

Figure \ref{fig:resextrapol} shows the extrapolation of the point resolution in $r\varphi$ to a magnetic field of \SI{3.5}{\tesla} and a drift length of \SI{2.35}{\m}, as planned for the ILD detector. For this extrapolation, equation \eqref{math:zResolutionElectronLoss} has been used. The values for $\sigma_{0,r\varphi}$ and $N_{\mathrm{eff}}$ are taken from the fit to the measured resolution, see table~\ref{tab:res}. The transverse diffusion constant at \SI{3.5}{\tesla} is derived using a Magboltz simulation to be $D_t=\SI[per-mode=symbol]{0.030}{\mm \per \sqrt{\cm}}$. The upper curve of the plot shows the result for the measured attachment rate of \SI{0.495}{\per \m}, the lower curve the extrapolated point resolution under the assumption of an attachment rate of zero. The $1\sigma$ error bands are based on the errors on the fit of the parameters as listed in table~\ref{tab:res}. It is visible that the required point resolution of \SI{100}{\um} over the full drift length at the ILD TPC is possible if the 
gas quality 
is tightly 
controlled and impurities are minimised.

\section{Conclusion}
Readout modules based on a triple GEM stack were operated in a prototype TPC chamber. The performance of the system was studied in detail. 
Stable operation could be established. Significant field distortions were observed close to the edges of the modules, and alignment effects between neighbouring modules were studied. Based on data, both distortion and alignment effects were corrected for. The intrinsic point resolution of the system was measured to be close to \SI{70}{\um}, with an increase as a function of the drift distance compatible with diffusion effects. Based on these results, a time projection chamber using a GEM-based amplification scheme and a modular readout structure were shown to perform well and fulfil the requirements for an experiment at the International Linear Collider.

\section{Acknowledgements} 

This material is based upon work supported by the National Science Foundation under Grant No.~0935316 and was supported by JSPS KAKENHI Grant Number 23000002. The research leading to these results has received funding from the European Commission under the 6th Framework Programme ``Structuring the European Research Area'', contract number RII3-026126, and under the FP7 Research Infrastructures project AIDA, grant agreement no.~262025.\\
Special thanks go to Y.~Makida, M.~Kawai, K.~Kasami and O.~Araoka of the KEK IPNS cryogenic group, and A.~Yamamoto of the KEK cryogenic centre for their support in the configuration and installation of the superconducting PCMAG solenoid. \\
The measurements leading to these results have been performed at the Test Beam Facility at DESY Hamburg (Germany), a member of the Helmholtz Association.\\
The authors would like to thank the technical team at the DESY II accelerator and test beam facility for the smooth operation of the test beam and the support during the test beam campaign. The contributions to the experiment by the University of Lund, KEK, Nikhef and CEA are gratefully acknowledged.

\section{Disclaimer}
Any opinions, findings, and conclusions or recommendations expressed in this material are those of the author(s) and do not necessarily reflect the views of any of the funding agencies involved. No warranty expressed or implied is made with regard to any information or its use in this paper.

\bibliographystyle{utphys}
\bibliography{main}

\end{document}